\documentstyle[11pt,aaspp4]{article}      

\lefthead{The et al.}
\righthead{Nuclear Reactions Governing the Nucleosynthesis of $^{44}$Ti}

\begin{document}

\title{Nuclear Reactions Governing the Nucleosynthesis of $^{44}$Ti}

\author{The, L.-S., Clayton, D. D., Jin, L., \& Meyer, B. S.}
\affil{Department of Physics \& Astronomy, Clemson University,
       Clemson, SC 29634-1911 }

\begin{abstract}

Large excesses of $^{44}$Ca in certain presolar graphite and silicon
carbide grains give strong evidence for $^{44}$Ti production in
supernovae.  Furthermore, recent detection of the $^{44}$Ti
$\gamma$-line from the Cas A SNR by CGRO/COMPTEL shows that radioactive
$^{44}$Ti is produced in supernovae.  These make the $^{44}$Ti
abundance an observable diagnostic of supernovae.

Through use of a nuclear reaction network, we have systematically
varied reaction rates and groups of reaction rates to experimentally
identify those  that govern $^{44}$Ti abundance in core-collapse
supernova nucleosynthesis.  We survey the nuclear-rate dependence by
repeated calculations of the identical adiabatic expansion, with peak
temperature and density chosen to be 5.5$\times$10$^9$K and 10$^7$ g
cm$^{-3}$, respectively, to approximate the conditions in detailed
supernova models.  We find that, for equal total numbers of neutrons
and protons ($\eta$=0), $^{44}$Ti production is most sensitive
to the following reaction rates:  $^{44}$Ti($\alpha$,p)$^{47}$V,
$\alpha$(2$\alpha$,$\gamma$)$^{12}$C, 
$^{44}$Ti($\alpha$,$\gamma$)$^{48}$Cr, $^{45}$V(p,$\gamma$)$^{46}$Cr.
We tabulate the most sensitive reactions in order of their importance
to the $^{44}$Ti production near the standard values of currently
accepted cross-sections,
at both reduced reaction rate (0.01$\times$) and at 
increased reaction rate (100$\times$) relative to their standard values.
Although most reactions retain their importance for $\eta >$ 0, that
of $^{45}$V(p,$\gamma$)$^{46}$Cr drops rapidly for $\eta\geq$0.0004.
Other reactions assume greater significance at greater neutron excess:
$^{12}$C($\alpha$,$\gamma$)$^{16}$O,
$^{40}$Ca($\alpha$,$\gamma$)$^{44}$Ti,
$^{27}$Al($\alpha$,n)$^{30}$P, $^{30}$Si($\alpha$,n)$^{33}$S.
Because many of these rates are unknown experimentally, our results
suggest the most important targets for future cross section
measurements governing the value of this observable abundance.

\end{abstract}

\keywords{nuclear reactions, nucleosynthesis, abundances}

\newpage \section{INTRODUCTION}

Radioactive $^{44}$Ti, produced in core collapse supernovae, is an
isotope of extraordinary astrophysical significance.  Its primary
observable effects appear to be three in number and will likely grow in
coming years.  The first is that the relatively large abundance of
$^{44}$Ca--it is the second most abundant calcium isotope and the 44th
most abundant species overall in solar system material--is
overwhelmingly due to its synthesis as $^{44}$Ti parent (Bodansky,
Clayton, \& Fowler 1968; Woosley, Arnett, \& Clayton 1973;
\cite{tim96}).  The second is that resulting gamma rays from young core
collapse supernovae are expected to be visible from several Galactic
remnants (\cite{cla71}; \cite{har93}; \cite{lei94}; \cite{dup97}).
$^{44}$Ti decays to $^{44}$Sc emitting 67.9 keV (100\%) and 78.4 keV
(98\%) lines. $^{44}$Sc then decays into $^{44}$Ca, which emits a 1.157
MeV (100\%) line.  The search for Galactic $^{44}$Ti gamma-ray lines
has been carried out by both the gamma-ray spectroscopy experiments on
HEAO 3 (\cite{mah92}), the SMM satellite (\cite{lei94}), and by CGRO
surveys with COMPTEL (\cite{dup97}).  Only from the direction of Cas A
SNR has the 1.157 MeV $^{44}$Ti gamma line flux been detected by
COMPTEL (\cite{can78}; \cite{the95}), with a flux initially reported at
(7.0$\pm$1.7)$\times$10$^{-5}$ $\gamma \;$ cm$^{-2}\;$ s$^{-1}$.
Subsequently Iyudin et al. (1997) have reported CGRO cycle 1-5 results
from a 2$\times$10$^6$ s exposure of Cas A.  The result is a flux of
(4.8$\pm$0.9)$\times$10$^{-5}$ $\gamma \;$ cm$^{-2}\;$ s$^{-1}$ and
significance of detection of 6$\sigma$.  This measurement is reasonably
consistent with the CGRO/OSSE three-line flux measurements (The et al.
1996) of (1.8$\pm$1.5)$\times$10$^{-5}$ $\gamma \;$ cm$^{-2}\;$
s$^{-1}$ and with the preliminary result of RXTE/HEXTE AO1 \& AO2
observations (\cite{rot98}) of (1.3$\pm$1.2)$\times$10$^{-5}$ $\gamma
\;$ cm$^{-2}\;$ s$^{-1}$, at 4\% confidence level with the most
probable flux of the combined three instrument measurements being 3.2
$\times$10$^{-5}$ $\gamma \;$ cm$^{-2}\;$ s$^{-1}$.  This flux
translates into (1.3, 7.8)$\times$10$^{-4}$ M$_{\sun}$ of $^{44}$Ti for
a $^{44}$Ti half-life of (66.6, 39.0) yrs respectively (\cite{alb90};
\cite{mei96}), distance of 3.4 kpc to Cas A (\cite{ree95}) and age of
315 yrs (\cite{ash80}).  Because the $^{44}$Ti yield probes the
dynamics of core collapse supernova nucleosynthesis, and in particular,
the location of the mass cut, the pre-supernova composition inside
$\sim$2 M$_{\sun}$, and the maximum temperature and density reached
during the passage of the shock wave in the ejecta, this detection has
generated great enthusiasm for SNR distance measurements
(\cite{ree95}), for $\gamma$-ray line observations (\cite{iyu97};
\cite{rot98}; \cite{the96}), for new nuclear laboratory experiments to
accurately determine the $^{44}$Ti half-life 
(\cite{alb90}; \cite{mei96}; \cite{nor97}; \cite{gor98}), 
and for theoretical nucleosynthesis calculations in the context of 
core-collapse supernovae (\cite{woo91}; \cite{tim96}; \cite{nag97}).

The third effect is that $^{44}$Ca-enriched silicon-carbide particles
extracted from meteorites have been identified (\cite{ama92};
\cite{Nit96}; \cite{hop96}) as presolar particles that condensed within
supernova ejecta during their first few years of expansion, while
$^{44}$Ti was still at its initial value (\cite{cla75}).  These grains
are of enormous value in probing the dynamics and make up of supernova
ejecta (\cite{caz97}).

In the near future, we can expect very accurate $^{44}$Ti line flux
measurements, precise distances to several young SNRs, and a accurate
$^{44}$Ti half-life (recent measurements of $^{44}$Ti lifetime
by \cite{nor97} and \cite{gor98} seem to converge at a value of 
(87.7$\pm$1.7) y).  
This reduction of current uncertainties will
allow more meaningful comparisons of the $^{44}$Ti mass in Cas A and
other young SNRs (in particular, Tycho and Kepler) to supernova
models.  However, a remaining factor limits such comparisons between
observations of $^{44}$Ti mass and supernova models, namely, the
uncertainty in the nuclear cross sections governing the synthesis of
$^{44}$Ti.  A large number of nuclear reactions play a role in
$^{44}$Ti production, and the cross sections for many of these are only
estimated from nuclear models.  While these estimated cross sections
should be fairly accurate in many cases (\cite{rauscher}), there is no
guarantee that they are so, and only laboratory measurements will
provide us with confidence in the yield predictions from
nucleosynthesis calculations.  Cross section values enter linearly into
the stellar reaction rate.  For this reason, and because of $^{44}$Ti's
astrophysical significance, we have surveyed the relative importance of
the nuclear reaction rates that govern its abundance. 
Our purpose is to
identify for the nuclear physics community those reaction rates that
have the greatest astrophysical importance to the $^{44}$Ti abundance
and which, consequently, would be appropriate targets for future cross
section measurements.
Using a nuclear reaction network with 378 isotopes, 
we have systematically varied specific reaction rates and
groups of reaction rates in an experimental test for astrophysical
significance.
We identify the most important nuclear rates and explain
our understanding of the overall rate dependencies we have found.
Preliminary results were presented in Jin et al. (1997).  
In the following we will refer to the terms ``cross section'' and
``reaction rate'' somewhat interchangeably. It should be understood
that ``reaction rates'' are calculated from either theoretically
predicted and/or experimentally measured cross sections.
Related to this survey is work done by Woosley \& Hoffman (1991) in which they
varied the peak temperatures, densities, and neutron excesses of
explosive silicon burning in alpha-rich freezeout to find the range of
$^{57}$Co and $^{44}$Ti production with standard reaction rates.

\section{Overall Strategy}

\subsection{Laboratory Nuclear Astrophysics}

The present work is part of a long-range effort to determine the
identity of the nuclear reactions that are of greatest importance to
nucleosynthesis in stars.  The goal is to improve the accuracy of
nucleosynthesis calculations by encouraging experimental study of key
reactions. Although this has always been an important part of the
science, as embodied for example in the Nobel-Prize winning work of 
W. A. Fowler, it has not always been emphasized with useful clarity. 
Most nucleosynthesis calculations have been concerned with testing the
fundamental paradigm: does the evolution and explosion of stars
reproduce the known elemental and isotopic abundances?  Most research
efforts address more explicitly the astrophysical models than the
nuclear data base. That data base is usually taken as a given, and so
most calculations have used the tabulated values as given in order to
calculate the nuclear production within specific stellar models. All
have been aware in this that the nuclear data base is uncertain, and
conclusions have been drawn that respect that uncertainty.  But almost
no nucleosynthesis calculations display the dependence of the
abundances on the values of the reaction rates used.  The community of
laboratory nuclear physicists devoted to this problem has continued the
work of improving that data base. Indeed the level of interest in this
has been increasing in the past few years, in part because of the
construction of several new radioactive-ion-beam facilities 
(\cite{lub97}), each of which has realized that those
facilities offer new opportunities to clarify cross sections near the
Z=N line in the nuclide chart where several important astrophysical
processes occur.

What is often missing is a clear idea of the relative importance of
specific reactions for the overall problem. Laboratory scientists quite
understandably do not wish to pursue all of the large numbers of reactions
that naturally occur. It has always been the demonstrated importance of
the exact value for a cross section that has inspired improved
measurements of it. This has usually been rather clear in those basic
cycles instrumental for nuclear power in stars, where the identity of
the important reactions is visible.  The important reactions during
advanced burning processes responsible for many of the intermediate
mass nuclei have not been so closely scrutinized, however. During these
processes a large number of reactions occur simultaneously, often
presenting parallel paths to the same result, so that the significance
of the value of each is obscured.  The size of a cross section is no
sure guide to the importance of the value attached to it.  In certain
circumstances during advanced burning a huge reaction rate is often 
opposed by a huge and nearly equal flow due to the inverse reaction, 
nullifying importance for the exact value of that reaction rate for 
that epoch of that process. So complex are those networks that unambiguous
responsibility for final abundances has been difficult to assign.  
This is especially true for nuclear processes taking place in or near
equilibrium.
Our approach to clarifying this is to explicitly change the value of
a specific reaction rate (or rates) 
and to redo the calculation to test the sensitivity of a final answer
to its value. We have developed a logical sequence to do this in a
practical manner.

\subsection{Four Requirements for Meaningful Measurements}

Because the goal is to ultimately influence measurements of key cross
sections, it is useful to list explicitly the four requirements that
are, at a minimum, needed for this goal to be achieved:
\begin{enumerate} 
\item An appropriate astrophysical model of events significant for
      nucleosynthesis; 
\item An observable from that process, usually an abundance result 
      that is either known or measurable; \item Dependency of the value
      of the observable on the value of a nuclear cross section;
\item Experimental strategy for measuring that cross section, or at
      least of using measurable data to better calculate it.
\end{enumerate} 
It is important to note in item 1 that the model must
be ``appropriate'', not necessarily ``correct''. We know that models of
supernovae, for example, are never ``correct''. The unknowns from all
of physics are many. But the model may be appropriate for testing the
sensitivity of the observable to the value of the cross section in
question. The cross-section dependence may be determinable with more
accuracy than the absolute value of the observable, because the
absolute value depends upon the model's realism.  Indeed, the model's
realism will later be tested by the nucleosynthesis results it produces
once the important reaction rates are known. We stress this distinction
that has confused many who ask, ``Why is it so important to pin down
the key cross sections when the astrophysical model introduces more
uncertainty than the cross sections do?''

The observable will usually be some abundance in the natural world.  It
may be the bulk abundance in solar system matter. It may be the
abundance outside a supernova that has just created new nuclei. It may
be an isotopic anomaly within presolar grains having isotopic
composition distinct from solar isotopes. It could be abundances within
cosmic rays. Other examples exist. The point is that the observable
should on good independent grounds be believed to measure the process
modeled. Combination of 1 and 2 focuses on the cross section dependence
of the value of an observable as produced in the process 1.  It is then
a separate question as to whether that cross section has a different
importance for another observable, or within a different process. Such
distinctions are necessary and valid.

The greater the sensitivity of the observable to the value of a specific 
reaction rate, the more important that rate's value is. Not every
abundance observable depends sensitively on the nuclear reaction 
rates;  for such cases a cross section does not carry significance
(for that observable and that process). When experiments are difficult
and costly it would be discouraging to learn afterwards that the effort
was without astrophysical significance. Our approach is to explicitly
vary the reaction rate to measure the dependence of the final abundance
of the observable to the value of that rate. By that procedure
a quantitative measure of its significance can be presented.

Item 4, the strategy for the measurement, will be the issue decided by
experimental teams motivated to nuclear research for purposes of
nuclear astrophysics.

\section{The $^{44}$Ti Abundance in the Alpha-Rich Freezeout}

The sensitivity of $^{44}$Ti synthesis to variations in unmeasured
cross sections must be determined for the process that produces it.
Point 1 of \S2.2 requires an
appropriate model for that process, which is now known to be the
alpha-rich freezeout of matter that was initially in nuclear
statistical equilibrium or quasi-statistical equilibrium but that
freezes out with excess alpha particles (\cite{woo73}; \cite{thi90};
\cite{thi96}).  Although $^{44}$Ti is also produced during silicon
burning (\cite{bcf68}) and within explosive helium burning as it may
occur within He caps atop Type Ia supernovae (\cite{Liv95};
\cite{woo86}), it is believed that the alpha-rich freezeout is
responsible for most of natural $^{44}$Ca and it is certain that it is
responsible for the $^{44}$Ti gamma lines detectable from young Type II
remnants. We therefore choose in this work to determine the sensitivity
of the $^{44}$Ti yield within that process.  It still remains to select
an appropriate model for that process.  Actual supernova models are
both uncertain and complicated.  In figure \ref{fig:Ti44contour} we
show contours of $^{44}$Ti yields computed with our network code
(described in more detail below) as functions of initial peak
temperature and density in nearly adiabatically expanding matter
experiencing an alpha-rich freezeout.  The figure shows that the
$^{44}$Ti yield variation is less than a factor of 10 for quite large
ranges of peak temperatures and densities near the reference
parameters, which have been suggested by hydrodynamic models.
Therefore, to survey the nuclear reaction rate sensitivity it will
suffice to evaluate that rate sensitivity within parameterized
expansions that are typical of the alpha-rich freezeout history within
the Type II core. We do this by adiabatic expansions of pure $^{28}$Si
matter initially at $T_9 = T/10^9$ K = 5.5 and mass density $\rho =
10^7$ g cm$^{-3}$.  These are the same conditions studied by Woosley \&
Hoffman (1991) for the alpha-rich freezeout (see their Table 1).  Such
conditions imply a ratio of the number density of photons to that of
nucleons of 0.57, corresponding to a specific entropy s/k $\approx 5$,
which is suitable for shock decomposed silicon in the Type II ejecta.
We take the initial density to decline exponentially with an e-fold
time 0.14 s, a typical hydrodynamical timescale, and we assume that the
photon-to-nucleon ratio stays fixed at 0.57 throughout the expansion;
thus, $\rho \propto T^3$.  This one zone calculation is then allowed to
cool until charged particle reactions freeze out ($T_9\sim$0.25).  Such
a schematic calculation, though inadequate to define the actual
$^{44}$Ti yield from core collapse supernovae, is adequate as the basis
for a survey of cross-section sensitivity for that yield.


One additional parameter set by the presupernova evolution is the 
fractional neutron excess $\eta$ of the matter undergoing the alpha-rich
freezeout. Roughly speaking its value, which depends upon how far 
thermonuclear evolution has progressed when the shock wave strikes, is
$\eta\simeq$0 in the He core, $\eta\simeq$0.002 ($Z_i/Z_{\sun}$) in the
CO core, and $\eta\simeq$0.006 in the Si core. We first survey at 
$\eta$=0 in section 4.

The reaction network used for the calculations is that described in
\cite{mey95} and Meyer, Krishnan, \& Clayton (1996).  
Table \ref{tab:tabnet} shows the
nuclides included in the network.  Within explosive silicon burning
that network includes many experimentally unknown reaction cross sections.  
For these reactions, the rates are calculated from the
SMOKER code (\cite{tat87}).


Figure \ref{fig:massfr} shows the evolution of mass fractions of key
species in the ``standard'' calculation.  The material begins as pure
$^{28}$Si, but quickly breaks down into light particles which then
establish a large quasi-equilibrium (QSE) cluster ranging from silicon
to beyond nickel.  For these particular conditions, the system does not
attain complete nuclear statistical equilibrium (NSE).  This is due to
the slowness of the triple-$\alpha$ reaction in maintaining the
abundance (per nucleon) $Y_h$ of heavy nuclei ($A \geq 12$) at the
level demanded by NSE.  This slowness imposes an extra constraint on
the equilibrium, which is thus a QSE, not an NSE (\cite{mkc97}).
Within the QSE cluster the rates of reactions involving $p$, $n$,
$\alpha$, and $\gamma$ reactions proceed at the same rate as their
inverses, just as in NSE; but in QSE the total number of nuclei in the
cluster is not the number they would have in NSE.


As the temperature falls, the $^{44}$Ti abundance drops.  This is a
consequence of the fact that the QSE cluster shifts upward in mass
owing to falling temperature in the face of a deficit of heavy nuclei
relative to NSE demands (\cite{mkc96}).  $^{44}$Ti breaks out of the
large QSE cluster around $T_9 = 4.3$, and its abundance starts to grow
as alpha particles in this alpha-rich freezeout reassemble into
$^{12}$C.  Subsequent alpha captures then carry the newly assembled
$^{12}$C up to higher-mass nuclei, including $^{44}$Ti.  The final
frozen $^{44}$Ti mass fraction ends up roughly three orders of
magnitude greater than its lowest value at the moment it broke out of
the large QSE cluster.
MPEG movies illustrating the nuclear dynamics of the ``standard''
calculation are available for viewing on the world-wide web at
http://photon.phys.clemson.edu/movies.html

\section{The Reaction-Rate Survey at $\eta$=0}

In order to determine the sensitivity of the $^{44}$Ti yield to
specific reaction rates, we carried out a systematic survey.
The central plan of the survey is to increase or decrease a reaction
rate or set of reaction rates by a large factor and then to rerun the
alpha-rich freezeout calculation described above.  The associated
errors in the nuclear cross sections may be either a uniform increase
or decrease of their values as a function of energy within the
statistical model, or they may be specific energy-dependent
contributions to the reaction rate.  We then compared the resulting
$^{44}$Ti yield to that in the ``standard'' calculation in which no
reaction rate was changed.  A significant yield difference indicates a
sensitivity to the varied rates.  It is important to note that the
network code automatically computes reverse reaction rates from the
(largely known) nuclear masses, forward reaction rates, and detailed
balance; thus, for example, if the rate for the reaction $^{48}{\rm
Cr}(\alpha, \gamma)^{52}{\rm Fe}$ is increased by a factor of 100, the
code automatically increased the rate for the reverse reaction
$^{52}{\rm Fe}(\gamma, \alpha)^{48}{\rm Cr}$ by a factor of 100.  This
is in compliance with detailed balance and is necessary to allow the
network to relax to the appropriate equilibrium when relevant.

We began with a survey over element number $Z$.  This step multiplies
all charge-increasing and mass-increasing
reaction rates on all isotopes of a single element $Z$ by a factor
of 0.01 and ran the alpha-rich freezeout.  This set of reaction consists
of ($\alpha$,$\gamma$), ($\alpha$,p), ($\alpha$,n), (p,$\gamma$),
(p,n), (n,$\gamma$).  Elements from carbon ($Z=6$)
to bromine ($Z=35$) are tested in this way.  We then repeated the
procedure, but this time using a factor of 100.  The results are shown
in figure \ref{fig:zsurv}.  From this figure it is evident that the
$^{44}$Ti yield is sensitive to reaction rates on a number of elements,
particularly on Ti itself and on V.  What is perhaps more striking,
however, especially in the lower panel, is the large number of elements
to which the $^{44}$Ti yield is {\it not} sensitive.  For example, the
$^{44}$Ti yield (the astrophysically relevant observable) is not
sensitive to large (factor of 100) uncertainties in reactions on any
isotope of carbon through phosphorus!  Such a negative result insofar
as rate sensitivity is concerned is as valuable as a positive result
because it increases the robustness of calculations of nucleosynthesis
in alpha-rich freezeouts.  The uncertainties in these cross sections do
not jeopardize the calculated $^{44}$Ti yields. Understanding this
prevents incorrect astrophysical motivations for nuclear experiments on
reactions irrelevant to the observable.


There is a simple reason for the insensitivity of the $^{44}$Ti yield
to reaction-rate variations on so many elements.  As discussed in the
previous section, the $^{44}$Ti abundance builds up as new heavy nuclei
assemble late in the expansion.  A nearly steady nuclear flow, whose
magnitude is governed by the triple-$\alpha$ rate, carries these new
nuclei up to higher mass.  By steady flow, we mean that the nuclear
flow into a nuclear species by capture from lower-mass nuclei equals
the capture flow out of that species; thus, the abundance of the
nuclear species in question does not change except over long time.
Because the flow is given by a rate times an abundance, an increase in
the rate is compensated by a decrease in the abundance so that the net
nuclear flow upward remains unchanged.  By analogy, if a smoothly
flowing river widens at some point, its depth or its speed decreases in
order to keep the same number of gallons per second moving downstream.
As a particular nuclear example, a factor of 100 decrease in the
$^{36}$Ar abundance compensates the factor of 100 increase in the
$^{36}{\rm Ar}(\alpha , \gamma) ^{40}{\rm Ca}$ reaction rate in a
steady flow.  A large uncertainty in the $^{36}{\rm Ar}(\alpha ,
\gamma) ^{40}{\rm Ca}$ rate would thus certainly be important for the
yield of $^{36}$Ar, but it is essentially irrelevant for our chosen
observable, the $^{44}$Ti yield.  These considerations illustrate the
importance of clearly defining the observable when discussing the
importance of particular nuclear reactions.


Table \ref{tab:tabzrank} ranks by element the $^{44}$Ti yield variation
resulting from the survey over each element.  The percent change is the
increase in $^{44}$Ti yield divided by the reference value, in parts
per hundred.  A large variation for a given element means that the
final $^{44}$Ti yield is sensitive to at least one reaction rate on
some isotope of that element.  The next task is to identify the key
isotope.  For each element showing a large $^{44}$Ti-yield variation,
we performed an isotope search by multiplying all reactions on a given
isotope of that element $Z$ by a factor of 0.01 and rerunning the
alpha-rich freezeout.  This was repeated for each isotope of the
element $Z$ in the network.  After completing the isotope survey for
element $Z$, we surveyed the next important element.  This entire
procedure was then repeated but with a factor 100 multiplying the
reaction rates.


Figure \ref{fig:asurv} shows the results for the isotope surveys for
Ar, Ca, Ti, and V for both the 0.01 and 100 factors.  It demonstrates
that the $^{44}$Ti yield significantly varies for changes on only one
isotope of each of the elements.  For the elements shown, these are
$^{36}$Ar, $^{40}$Ca, $^{44}$Ti, and $^{45}$V.  Table
\ref{tab:tabasurv} ranks the particular isotopes for the 0.01 and 100
reaction-rate variation surveys.  Comparing the first entries of tables
\ref{tab:tabzrank} and \ref{tab:tabasurv}, for example, shows that a
100-fold decrease of the reaction rates for all Ti isotopes produces an
almost identical increase (+373\%) in final $^{44}$Ti abundance, in
comparison to the increase (+372\%) for the $^{44}$Ti reaction rate only.


With the key isotopes now identified, we next determined which
particular nuclear reaction dominated the sensitivity of the $^{44}$Ti
yield.  This step of the survey, similar to that done above, varied a
specific reaction on the key isotope.  The first such survey was of the
nuclear reactions on $^{45}$V.  We multiplied the $^{45}{\rm
V}(n,\gamma)^{46} {\rm V}$ rate by a factor of 0.01 and reran the
alpha-rich freezeout.  We then repeated this for the $(p,\gamma)$,
$(p,n)$, $(\alpha,\gamma)$, $(\alpha,n)$, and $(\alpha,p)$ reactions.
We did this for each isotope listed in table \ref{tab:tabasurv}.  We
then repeated the procedure with a multiplication factor 100.


The results for the surveys on $^{40}$Ca, $^{44}$Ti, $^{45}$Ti, and
$^{54}$Fe are shown in figure \ref{fig:rsurv}.  Its discrete abscissa
is the reaction channel for that isotope.  The results clearly show
that the $^{44}$Ti yield is sensitive to variations in the rates of the
particular reactions $^{40}{\rm Ca}(\alpha,\gamma)^{44}$Ti, $^{44}{\rm
Ti} (\alpha,p)^{47}$V, $^{44}{\rm Ti}(\alpha,\gamma)^{48}$Cr,
$^{45}{\rm V}(p,\gamma)^{46}$Cr, and $^{54}{\rm Fe}
(\alpha,n)^{57}$Ni.  Table \ref{tab:rsurvlowhigh}  ranks the particular
nuclear reactions according to their influence at $\eta$=0 on the final 
$^{44}$Ti yield.  For example, reduction of $^{44}{\rm
Ti}(\alpha,\gamma)^{48}{\rm Cr}$ has almost no effect (unless of course
the $(\alpha,p)$ branch is also reduced), whereas increase of its rate
by 100$\times$ is almost as effective as an increase of the
$(\alpha,p)$ channel.
 Such conclusions depend on the relative magnitude of proton and gamma
widths of the compound nuclear states, and on the factor by which they
are altered.  Owing to its special role in the late assembly of heavy
nuclei, we also surveyed the triple-$\alpha$ reaction and include it in
table \ref{tab:rsurvlowhigh}.


One might have predicted that a number of these reactions would be
important, although the magnitude of their effects would not have been
as easy to guess.  The $^{45}{\rm V}(p,\gamma)^{46}$Cr reaction, the
one showing the largest effect, however, was not one for which we had
anticipated a special role.  This underscores the importance of making
a systematic survey when considering the reaction-rate sensitivity of a
particular observable.  In fact, not only are systematic surveys
important for clearly delineating reaction-rate sensitivities, but they
can also clarify nucleosynthesis.  It was only after developing new
insights into the nuclear dynamics of an alpha-rich freezeout that we
were able to understand the reason for the importance of the 
$^{45}$V(p,$\gamma$)$^{46}$Cr reaction.  
By the same token we can understand
why its importance vanishes for neutron richness $\eta>$0.0004 (see
Section 6).  We return to this issue in the next section.

The remaining task for the $^{44}$Ti survey was to determine the form
of the sensitivity of each particular important reaction.  Figure
\ref{fig:rsurv} shows that increasing the $^{45}{\rm
V}(p,\gamma)^{46}$Cr reaction by a factor of 100 decreases the
resulting $^{44}$Ti yield by a factor of nearly 50.  This suggests a
nearly linear dependence of the $^{44}$Ti yield on the rate of this
reaction.  For a decrease by a factor of 100 on this rate, however, the
$^{44}$Ti yield increases by less than a factor of two, suggesting a
non-linear dependence.  The yield dependence on the rate value
determines its significance.  To quantify the form of the dependence,
the alpha-rich freezeout was repeated many times, each time with a
specific important nuclear reaction rate multiplied by a factor between 0.01
and 100.  This maps out the sensitivity of the $^{44}$Ti yield to any
change by a factor between 0.01 and 100 on the important nuclear
reactions.  Figures \ref{fig:45pg}, \ref{fig:3alpha}, \ref{fig:44ag},
and \ref{fig:44ap} show results of such surveys.  In each of these
figures, a small solid square indicates the result using the
``standard'' value for that particular rate.


These figures illuminate more subtle aspects of the reaction rate
sensitivities.  Figure \ref{fig:45pg} shows that a factor of two
increase or decrease in the $^{45}{\rm V}(p,\gamma)^{46}$Cr rate
results in a $\sim 20\%$ change in the yield of $^{44}$Ti.  On the
other hand, figure \ref{fig:44ag} shows that a factor of two change in
the $^{44}{\rm Ti}(\alpha,\gamma)$ reaction rate barely changes the
$^{44}$Ti yield, even though a factor of 100 increase in the rate
dramatically drops the yield.  Because the reaction rates we use are
either experimentally known or among the best theoretical estimates
presently available, we expect the true rates in general not to be more
than, say, an order of magnitude different.  Thus, the quantity
expressing the $^{44}$Ti-sensitivity to a particular reaction near its
reference value is the slope of the curve for that reaction, in analogy
to figure \ref{fig:45pg}.  In table \ref{tab:tabslope} we present these
quantities, in a ranking by magnitude, for the sixteen most important
reactions at $\eta$=0 insofar as final $^{44}$Ti abundance is concerned. 


As an example of how to use table \ref{tab:tabslope}, consider the case
of $^{45}{\rm V}(p,\gamma)^{46}$Cr.  If it were measured, one would
then compare that rate (especially near $T_9=2$ for reasons to be
discussed in the next section) to the value used in our surveys, the
fits for which we present in table \ref{tab:ourrates}.  If the measured
rate were a factor of 1.5 greater than the tabulated value for all
temperatures, the resulting $^{44}$Ti yield would change by
$-0.361\times(1.5-1)=-0.18=-18$\%, that is, it would be 82\% of the
standard value.


It was only through a large and systematic survey that we were able to
determine the relative importance of the reactions listed in table
\ref{tab:tabslope}.  This was a computationally intensive effort
comprising more than 2000 sets of alpha-rich freezeout calculations,
each taking $\sim$20 minutes on a 200 MHz DEC Alpha workstation.
Nevertheless, it is only in this way that one can be fully confident of
understanding the reaction-rate sensitivity of the $^{44}$Ti yield.

 A final part of the survey applies to neutron-excess $\eta$ greater
than zero. The survey is repeated in summary form in Section 6 for
$\eta$ = 0.002 and $\eta$ = 0.006. For these neutron enrichments, 
$^{44}$Ti synthesis was unaffected by the value of the 
$^{45}$V(p,$\gamma$)$^{46}$Cr reaction rate,
a result also explained in Section 6. But first we describe the reasons
for the importance of the Table \ref{tab:rsurvlowhigh} reactions at 
$\eta$ = 0.

\section{Discussion of the Most Important Reactions}

Table \ref{tab:tabslope} lists the reactions to which the $^{44}{\rm
Ti}$ yield is most sensitive in an alpha-rich freezeout.  In this
section, we briefly explore the reasons for this sensitivity for three
important reactions.  In addition to the immediate interest of
understanding the importance of these reactions, this exercise provides
new insights into the nuclear dynamics of the alpha-rich freezeout.

Any nucleosynthetic system seeks to maximize its entropy and thereby to
achieve an equilibrium.  Its ability to do this, however, is governed
by the rate of certain key reactions.  When a particular reaction
becomes too slow, the equilibrium breaks.  The system still maximizes
the entropy, but the slowness of the particular reaction adds a
constraint which restricts the number of accessible states
(\cite{mkc97}).  As the system evolves toward lower temperatures, more
reactions become slow, more equilibria break, and more constraints are
added to the entropy maximum.  In this way, the system descends the
``hierarchy of statistical equilibria'' 
(contribution of Meyer in \cite{wal97}).  Several of the
reactions in table \ref{tab:tabslope} are important because they
control the breaking of equilibria and, thus, the descent of the
hierarchy.  Other reactions are important because they control when the
steady (but non-equilibrium) flow into $^{44}$Ti breaks down or because
they control the economy of light particles that influence the
abundance of $^{44}$Ti late in the expansion.

\subsection{The triple-$\alpha$ reaction}

We begin with the triple-$\alpha$ reaction.  Figure \ref{fig:triple-a}
shows the time history of the $^{44}$Ti mass fraction during two
alpha-rich freezeouts--one with the standard (\cite{CF88}) rate (dashed
curve) and one with the standard rate increased by a factor of 100
(solid curve).  For comparison, the dotted curve gives the $^{44}$Ti
mass fraction in NSE.  This NSE is calculated at each timestep in the
expansion using the same temperature, density, and neutron richness as
in the network.  The dashed and solid curves differ throughout the
expansions.  In the case with the increased triple-$\alpha$ rate, the
nuclear system achieves and maintains NSE for the early part of the
expansion.  With the standard triple-$\alpha$ rate, the system never
reaches NSE but rather only a QSE in which the number of heavy nuclei
$Y_h$ differs from that in NSE.  Below $T_9 \approx 5.3$ the number of
heavy nuclei in the standard run QSE is less than that in NSE, so the
heavy nuclei are in the presence of an overabundance (relative to NSE)
of light particles.  This favors nuclei heavier than $^{44}$Ti (e.g.
$^{56}$Ni), so the $^{44}$Ti mass fraction is far below that of NSE
until later in the expansion when the large QSE breaks and the
$^{44}$Ti can build back up without being driven upward to $^{56}$Ni.


The $^{44}$Ti mass fraction in the standard calculation is less than
that in the case with the increased triple-$\alpha$ reaction for $T_9 <
5.3$.  The latter expansion deviates from NSE beginning at $T_9 \approx
5.2$.  At this point the nuclear system is also in a QSE, but this QSE
has a greater number of heavy nuclei and a smaller number of light
particles than the standard case QSE.  This yields a greater mass
fraction of $^{44}$Ti within the QSE cluster.  At the point $^{44}$Ti
breaks out of the large QSE cluster containing $^{56}$Ni (the dip in
both curves near $T_9 = 4$), the $^{44}$Ti mass fraction in the
expansion with the increased triple-$\alpha$ rate exceeds that in the
standard case by a factor of $\sim$ eight.  More importantly, the
expansion with the increased triple-$\alpha$ rate has a factor of $\sim
4.6$ lower mass fraction of $^4$He.  Remarkably, this factor of 4.6
(when cubed) compensates for the increased triple-$\alpha$ cross
section; thus, the two expansions create new nuclei at the nearly the
same rate at late times.  The crucial aspect for $^{44}$Ti is that the
lower $^4$He abundance in the expansion with the increased
triple-$\alpha$ cross section requires larger abundances of nuclei
between $^{12}$C and $^{56}$Ni to carry the same flow.  This causes the
increase in the final yield of $^{44}$Ti.

\subsection{$^{44}{\rm Ti}(\alpha,p)^{47}{\rm V}$}

Figure \ref{fig:ti44ap} shows the $^{44}$Ti mass fraction in the
standard calculation (dashed curve) and in an expansion with the
$^{44}{\rm Ti} (\alpha , p)^{47}{\rm V}$ rate increased by a factor of
100.  Also shown as the dotted curve is the QSE mass fraction of
$^{44}$Ti in the standard expansion calculated at each timestep from
the same temperature, density, neutron richness, and number of heavy
nuclei as in the network.  The evolution of the $^{44}$Ti mass fraction
follows QSE in both cases down to $T_9 \approx 4.5$.  The increased
$^{44}{\rm Ti}(\alpha, p)^{47}{\rm V}$ rate, however, allows the
$^{44}$Ti to remain in QSE with $^{56}$Ni longer; thus, when the QSE
finally breaks, the $^{44}$Ti mass fraction is $\sim 30$ times lower
than the corresponding value in the standard case.  This result
illustrates the important role $^{44}{\rm Ti}(\alpha,p)^{47}{\rm V}$
plays in linking the Si-Ca QSE cluster to the Ni-centered QSE cluster
(\cite{woo73}; \cite{hix96}).  Once the two clusters fall out of
equilibrium with each other, the $^{44}{\rm Ti}(\alpha,p)^{47}{\rm V}$
reaction governs the rate at which abundance moves from the Si-Ca-Ti
cluster to the Ni cluster.  A faster rate gives a lower abundance in
the Si-Ca-Ti cluster and a lower $^{44}$Ti yield.


\subsection{$^{45}$V(p,$\gamma$)$^{46}$Cr}

Perhaps the biggest surprise in our survey was the importance of the
$^{45}{\rm V}(p,\gamma)^{46}{\rm Cr}$ reaction.  Figure \ref{fig:v45pg}
shows the $^{44}$Ti mass fraction for the standard calculation (dashed
curve) and for an expansion with the $^{45}{\rm V} (p,\gamma)^{46}{\rm
Cr}$ increased by a factor of 100.  The $^{44}$Ti evolution is
precisely the same in the two expansions down to $T_9 \approx 2$.  The
subsequent deviation between the two cases, however, is dramatic.
While the $^{44}$Ti mass fraction drops by $\sim 30\%$ as $T_9$ falls
below 2 in the standard calculation, it plummets by a factor of $\sim
100$ for the increased $^{45}{\rm V}(p,\gamma)^{46}{\rm Cr}$ rate.  The
reason is, as in the previous cases, that this reaction controls the
breaking of an equilibrium.


The equilibrium of importance in this case is a $(p,\gamma)-(\gamma,p)$
equilibrium among the $N=22$ isotones.  In such equilibrium, the rates
(as opposed to the cross sections) of (p,$\gamma$) reactions proceed at
the same rates as the reverse ($\gamma$,p) reactions; i.e. $^{A}$Z + p
$\rightleftharpoons$ $^{A+1}$(Z+1) + $\gamma$, where the reversed
arrows indicate equal numbers of reactions per unit time. We focus on
the N = 22 isotones, A = Z + 22.  Figure \ref{fig:pgequil} shows how
this equilibrium breaks for the standard set of cross sections.  It
gives for four temperatures the ratio of the network abundance of these
isotones to the values they would have in $(p,\gamma)-(\gamma,p)$
equilibrium.  For $T_9=2.66$, all of the isotones for Ti through Fe are
in equilibrium with each other under exchange of protons, but the
lower-charge isotones have already broken out of this equilibrium.  As
the temperature falls to $T_9 = 2.15$, $^{48}$Fe strongly falls out of
equilibrium with the other four species, but also evident is the fact
that the two small equilibrium clusters ($^{44}$Ti \& $^{45}$V) and
($^{46}$Cr \& $^{47}$Mn) are starting to fall out of equilibrium with
each other; that is, $^{45}$V and $^{46}$Cr are no longer in
$(p,\gamma)-(\gamma,p)$ equilibrium with each other.  This break grows
with decreasing temperature.  By $T_9 = 1.81$ the two small
$(p,\gamma)-(\gamma,p)$ clusters, ($^{44}$Ti \& $^{45}$V) and
($^{46}$Cr \& $^{47}$Mn), are strongly out of equilibrium with each
other.  When this occurs, $^{45}{\rm V}(p, \gamma)^{46}{\rm Cr}$
produces net destruction of $^{45}$V, and hence of $^{44}$Ti, with
which it is in equilibrium.


The $T_9 = 1.81$ panel of this figure shows that the $^{44}$Ti-$^{45}$V
cluster is overabundant and the $^{46}$Cr-$^{47}$Mn cluster is
underabundant with respect to the overall $(p,\gamma)-(\gamma,p)$
equilibrium.  This indicates that too few nuclei have moved from the
$^{44}$Ti-$^{45}$V cluster up to the $^{46}$Cr-$^{47}$Mn cluster
because of the slowness of the reaction that links them, namely,
$^{45}{\rm V}(p,\gamma)^{46}$Cr.  By increasing the rate of this
reaction, the two small clusters remain in equilibrium with each other
longer, the $N=22$ isotones shift to higher mass, and more $^{44}$Ti is
destroyed.

The dotted curve in figure \ref{fig:v45pg} amplifies this point by
showing what would happen in the standard case to the $^{44}$Ti mass
fraction if the $(p,\gamma)-(\gamma,p)$ equilibrium among $N=22$
isotones persisted to low temperature.  The faster $^{45}{\rm V}
(p,\gamma)^{46}$Cr reaction allows the $^{44}$Ti to remain in the
$(p,\gamma)-(\gamma,p)$ equilibrium to lower temperature.  Because the
$(p,\gamma)-(\gamma,p)$ equilibrium abundances shift to higher mass
with lower temperature, the $^{44}$Ti mass fraction falls.  Even with
the fast $^{45}{\rm V}(p,\gamma)^{46}$Cr reaction, however, the
equilibrium eventually breaks and the $^{44}$Ti mass fraction freezes
out. In the surveys at greater neutron excess to follow, we demonstrate
that the controlling significance of $^{45}$V(p,$\gamma$)$^{46}$Cr
vanishes for $\eta >$ 0.0004. This will be understood as a consequence
of the smaller free proton density with increased $\eta$.

\section{Surveys Having Excess Neutrons}

We repeated the surveys in exactly the same procedure for a
nuclear gas containing more neutrons than protons in total. This initial
neutron richness, when the shock wave strikes the overlying matter, is not
altered during the subsequent rapid nuclear burning, which occurs too
rapidly for significant electron capture to occur. Thus the gas can be
described by a constant parameter 
$\eta = \sum_i (N_i-Z_i) Y_i$, where $N_i$ and $Z_i$ are
the mass fractions of nucleons (both bound in species $i$ and free)
and $Y_i \equiv X_i/A_i$ is the abundance of species $i$ with mass 
fractions $X_i$ and atomic mass number $A_i$.
The survey shows that although most nuclear reactions retain
their importance for $^{44}$Ti production during alpha-rich freezeouts, the
$^{45}$V(p,$\gamma$)$^{46}$Cr reaction is very sensitive to the value of 
$\eta$.  Demonstrating and understanding these facts is the main goal of this
section.

The range of anticipated values for $\eta$ is set by the presupernova
evolution and structure and by the strength of the shock wave. The
latter determines how far radially the shock can propagate with
sufficient strength to establish an alpha-rich freezeout, without which
the yield of $^{44}$Ti is too small to contribute substantially to the
natural $^{44}$Ca abundance. Within the entire He core of massive stars the
matter is dominated by He, $^{14}$N, $^{16}$O and $^{12}$C,
all of which have $\eta=0$.
But within the CO core $\eta$ has been increased by the conversion of $^{14}$N
to $^{18}$O and $^{22}$Ne, both of which have two excess neutrons. At solar
initial metallicity ($Z_\odot$) these comprise about 2\% by mass, so that
$\eta = 0.002$ throughout the CO core. Supernovae of lower initial metallicity
$Z_i$, such as supernova 1987A in the Magellanic Cloud, will have $\eta$ near
$0.002(Z_i/Z_{\odot})$ in their CO cores. Hydrostatic carbon burning does not
much increase $\eta$ within the NeMg core, but hydrostatic O burning does
do so. The more central core in which O has burned to create a Si core
has acquired more excess neutrons by electron captures on the products
of O burning (\cite{woo72}), yielding
$\eta$ in the range 0.003-0.006. In the interests of limiting the
computational demands, we have performed the full survey at the values
$\eta$ = 0.002 and 0.006. Finer precision mapping in $\eta$ was also done for
the $^{45}$V(p,$\gamma$)$^{46}$Cr reaction alone, after determining its
sensitivity to the neutron excess.

Tables \ref{tab:rsurvlowhigh2} and \ref{tab:rsurvlowhigh6} list 
the effect of reaction rate changes for the ten most
important regulators of $^{44}$Ti production at these two values of $\eta$.
Many of the same reactions remain important, such as 
$^{44}$Ti($\alpha$,p)$^{47}$V, $^{40}$Ca($\alpha$,$\gamma$)$^{44}$Ti,
and $^{36}$Ar($\alpha$,p)$^{39}$K; 
but the importance of $^{45}$V(p,$\gamma$)$^{46}$Cr disappears and some
new important reactions appear (e.g. $^{12}$C($\alpha$,$\gamma$)$^{16}$O).
More complete lists can be read
and downloaded from the Clemson nuclear astrophysics web site
(http://photon.phys.clemson.edu/tables.html).


In the interest of understanding we comment upon two of these dramatic
changes. The importance of the $^{45}$V(p,$\gamma$)$^{46}$Cr reaction declines
precipitously for $\eta >$ 0.0001, as shown in figure \ref{fig:V45pgeta}.
The cause is the decline of the
free-proton abundance $Y_p$ at every temperature with increasing $\eta$.
As $\eta$ increases, protons tend to be more tightly bound in nuclei.
Although the $^{44}$Ti-$^{45}$V pair remain in $(p,\gamma)$
equilibrium down to $T_9 {_\sim^<}$ 2 for $\eta$ as large as $\sim 0.01$,
the $^{45}$V abundance is declining with increasing $\eta$ due to the
diminishing free proton abundance.
As a result the final proton captures are unable to reduce the abundance of
the $^{44}$Ti-$^{45}$V pair during the freezeout.
Figure \ref{fig:V45pgeta}  shows the free proton abundance at $T_9=2$, 
the temperature at which the $(p,\gamma)$ equilibrium begins its freezeout, 
for several calculations at different $\eta$'s.  
In the equilibrium, the abundance ratio
$^{45}$V/$^{44}$Ti $\propto Y_p$.  The $^{45}$V(p,$\gamma$)$^{46}$Cr 
reaction rate is also proportional to $Y_p$; thus, the importance of 
this reaction rate in modifying the final $^{44}$Ti yield goes
as $Y_p^2$.  The fact that $Y_p$ at $T_9$ =2 drops off so rapidly for
$\eta > 0.0001$ explains why the final $^{44}$Ti yield sensitivity 
also declines so quickly.  The consequence is
that the $^{45}$V(p,$\gamma$)$^{46}$Cr reaction loses considerable
importance for observations of $^{44}$Ti in supernovae, at least unless they
have very small initial metallicity, or unless the shock wave is able to
reach the He shell, or unless some other process keeps the proton 
abundance higher
than would be expected simply from the alpha-rich freezeout.
It is worth noting that some Type Ia events create abundant $^{44}$Ti
(\cite{cla97});
but the burning is not literally an alpha-rich freezeout although it
does have hot excess alpha particles.  A variety of $\eta$'s might be expected
in this scenario depending on the initial composition.  For low $\eta$ we
will again expect certain proton capture reactions to affect the $^{44}$Ti
yield.  We will study the important cross sections for many observables 
in this setting in a subsequent work in this series.

The $^{12}$C($\alpha$,$\gamma$)$^{16}$O reaction assumes greater
importance for $\eta >0$.  The greater neutron richness of the matter causes
alpha particles to bind more tightly into nuclei, thereby decreasing the
alpha particle abundance throughout most of the expansion.  Because of the
smaller alpha particle abundance, flow of newly assembled nuclei to higher
mass freezes out at higher temperature.  When the cross section for
alpha particle capture on $^{12}$C is decreased, this freezeout occurs even
earlier, leading to considerably reduced $^{44}$Ti production.

\section{Conclusion}

We have analyzed the influence of 
individual reaction rates on $^{44}$Ti production in alpha-rich
freezeouts within Type II supernova events. Using direct surveys we
have established that the $^{44}$Ti production at $\eta$ = 0
is most sensitive to the following reaction rates:  
$^{45}$V(p,$\gamma$)$^{46}$Cr,
$\alpha$(2$\alpha$,$\gamma$)$^{12}$C, 
$^{44}$Ti($\alpha$,p)$^{47}$V,
$^{44}$Ti($\alpha$,$\gamma$)$^{48}$Cr,
$^{40}$Ca($\alpha$,$\gamma$)$^{44}$Ti, 
and 
$^{57}$Co(p,n)$^{57}$Ni.
For $\eta >$ 0 the two proton-induced reactions decline rapidly
in importance and some other reactions become crucial:
$^{12}$C($\alpha$,$\gamma$)$^{16}$O,
$^{40}$Ca($\alpha$,$\gamma$)$^{44}$Ti,
$^{27}$Al($\alpha$,n)$^{30}$P, and $^{30}$Si($\alpha$,n)$^{33}$S.
Several good reasons justify this effort at a depth of detail and
computational time that exceeds that normally devoted to nuclear
studies in astrophysics. Three main reasons are: (1) to provide
guidelines for laboratory measurements; (2) to establish better
diagnostics of supernova events; (3) to gain insight into complicated
and nonlinear nuclear reaction networks of importance for
nucleosynthesis.

The first reason parallels the large increase in recent activity in
laboratory nuclear astrophysics. In particular, many new
radioactive-ion-beam facilities are rapidly establishing the capability 
to measure reaction cross sections near and on the proton-rich side of
the $Z=N$ line. This capability opens to experimental study
nucleosynthesis processes that have heretofore had to rely on computed
cross sections. Each such experiment will be costly and time consuming,
however, so that solid grounds for the importance of any specific
reaction are welcome to experimental planners.

The second reason parallels the large increase in recent observations
of supernova nucleosynthesis. Although the mass of $^{44}$Ti within the
Cas A supernova 310-year-old remnant cannot yet be regarded as measured
with good precision, the recent detection of it insures that it
eventually will be well measured. When that precision has been
achieved, the mass of $^{44}$Ti will reflect the alpha-rich-freezeout
mass that was ejected. Since this mass depends in exciting ways on the
physics that underlies the Type II event (the mass cut and
the shock mechanism), all manner of exciting issues may hinge upon its
production.  The same will be true for the several recent Galactic
supernovae that will be detected with the improved gamma-ray detector
sensitivity that now exists in laboratories. It is probable that
significant variations of the $^{44}$Ti yield exist between different
events. Once that observational goal is achieved, the most accurate
cross sections will be wanted so that the scientific uncertainties
reflect the uncertainties over the realistic supernova model rather
than over the nuclear cross sections. Similar excitement attends the
measurement of $^{44}$Ti/$^{48}$Ti production ratios in samples of
supernova matter as recorded by supernova condensates found in
meteorites. One does not want nuclear cross section uncertainties to
stand in the way of quantitative analysis.

The third reason parallels the large recent increase in the
sophistication with which nucleosynthesis is viewed. The present study
illuminates this very well. Within the complex networks of explosive
nucleosynthesis, one can ideally think of the importance of any
specific nuclear reaction in various ways. It may have almost no
significance if that reaction maintains a partial QSE throughout the
burning. When QSE breaks down or fragments, a reaction may play a
rather complicated role in governing the changing relative total
numbers within the different QSE clusters; or it may play a role in the
density of free light particles; or it may simply be the specific
channel by which an abundance is altered in the freezeout. The
important reactions identified by us for $^{44}$Ti production in the
alpha-rich freezeout reveal each of these aspects. We have discussed
these to some extent in the discussions of the important reactions.

Finally we would observe that our study concentrated on one very
important nucleus within one nucleosynthesis process.  There are many
nucleosynthesis processes producing many different nuclei, so
additional studies of the present type are recommended. This requires
judgment as well as computing power.  Clearly one does not wish to
have huge numbers of irrelevant studies performed.  For this reason,
such studies should follow the ``four-requirement'' structure outlined
in \S 2.2.

\acknowledgments

We would like to thank the referee, Robert Hoffman for a thorough review
and suggestions to improve this paper.
This work has been supported by NASA grants NAG5-4329 and NAGW-3277.

\clearpage


\begin{deluxetable}{ccc} 
\footnotesize 
\tablecaption{Nuclei in the Network} 
\tablewidth{0pt} 
\tablehead{ 
\colhead{Proton Number, Z}  & \colhead{$A_{min}$}  & 
\colhead{$A_{max}$} 
}
\startdata 
1  & 2 & 3   \\ 
2  & 3 & 4   \nl 
3  & 6 & 8   \\ 
4  & 7 & 10  \\ 
5  & 8 & 11  \\ 
6  & 11 & 14 \\ 
7  & 12 & 15 \\ 
8  & 14 & 19 \\
9  & 16 & 21 \\ 
10 & 18 & 24 \\ 
11 & 19 & 27 \\ 
12 & 20 & 27 \\ 
13 & 22 & 30 \\ 
14 & 23 & 31 \\ 
15 & 27 & 34 \\ 
16 & 28 & 37 \\ 
17 & 31 & 40 \\ 
18 & 32 & 43 \\ 
19 & 35 & 48 \\ 
20 & 36 & 49 \\ 
21 & 40 & 49 \\ 
22 & 42 & 50 \\ 
23 & 44 & 50 \\    
24 & 44 & 50 \\     
25 & 46 & 59 \\ 
26 & 47 & 60 \\ 
27 & 50 & 63 \\ 
28 & 51 & 65 \\ 
29 & 57 & 70 \\ 
30 & 59 & 71 \\ 
31 & 59 & 79 \\
32 & 62 & 80 \\ 
33 & 65 & 85 \\ 
34 & 68 & 88 \\ 
35 & 69 & 91 \\
\enddata 
\label{tab:tabnet} 
\end{deluxetable}

\begin{deluxetable}{ccrccr} 
\tablecolumns{6} 
\tablewidth{0pc}
\tablecaption{ 
Order of importance of elements affecting the $^{44}$Ti
production in alpha-rich freezeout mechanism if the nuclear cross
sections for all mass or charge increasing reaction  
channels on all isotopes of an element are
changed by a factor of 1/100 or 100.  
} 
\tablehead{ 
\colhead{Rank} &
\multicolumn{2}{c}{Reaction rate} & \colhead{} &
\multicolumn{2}{c}{Reaction rate}  \\
     & \multicolumn{2}{c}{multiplied by 1/100} & &
       \multicolumn{2}{c}{multiplied by 100}  \\ \colhead{}  &
\colhead{Element}    & \colhead{\% $^{44}$Ti change} & \colhead{ } &
\colhead{Element} & \colhead{\% $^{44}$Ti change} } 
\startdata 
1  & Ti       & +372.6    &&  V   & -98.0 \\ 
2  & Ca       & -72.6     &&  Ti  & -91.3 \\ 
3  & V        & +56.6     &&  Co  & +30.6 \\ 
4  & Ni       & -52.3     &&  Ca  & +28.9 \\ 
5  & Ar       & -49.8     &&  Fe  & +9.2  \\ 
6  & Co       & -33.6     &&  Ni  & +7.1  \\ 
7  & S        & -20.8     &&  Ar  & +6.2  \\ 
8  & Cu       & -14.8     &&  Cr  & -5.9  \\ 
9  & N        & -14.8     &&  Cu  & +5.6  \\ 
10 & Si       & -4.5      &&  Na  & -2.4  \\ 
11 & Ne       & -4.0      &&  Mn  & +1.9  \\ 
12 & C        & +3.1      &&  C   & -1.8  \\ 
13 & Cr       & +2.9      &&  N   & -1.5  \\ 
14 & Na       & -1.7      &&  Mg  & +1.0  \\ 
15 & K        & -1.7      &&  S   & +0.8  \\ 
16 & Al       & +0.8      &&  Zn  & +0.5  \\ 
17 & Fe       & +0.7      &&  Ne  & +0.4  \\ 
18 & Mg       & -0.6      &&  Al  & -0.3  \\ 
19 & Mn       & -0.5      &&  Si  & +0.3  \\ 
20 & Zn       & -0.4      && P   & -0.2   \\ 
21 & O        & +0.4      &&  Cl  & +0.2  \\ 
22 & P        & +0.1      &&  K   & -0.1  \\ 
23 & Sc       & -0.1      && Sc  & +0.0   \\ 
\enddata 
\label{tab:tabzrank} 
\end{deluxetable}

\begin{deluxetable}{ccrccr} 
\tablecolumns{6} 
\tablewidth{0pc}
\tablecaption{ 
 The Order of Importance of Isotope to the $^{44}$Ti
 final mass fraction when all mass or charge increasing 
nuclear reactions of the isotope are
 multiplied by 1/100 or 100.  
} 
\tablehead{ 
\colhead{Rank} &
\multicolumn{2}{c}{Reaction rate} & \colhead{} &
\multicolumn{2}{c}{Reaction rate}  \\
     & \multicolumn{2}{c}{multiplied by 1/100} & &
       \multicolumn{2}{c}{multiplied by 100}  \\ 
\colhead{}  & \colhead{Isotope} & \colhead{\% $^{44}$Ti change} & \colhead{ } &
 \colhead{Isotope} & \colhead{\% $^{44}$Ti change} } 
\startdata 
1  & $^{44}$Ti  & +371.7 &&  $^{45}$V  & -98.0 \\ 
2  & $^{40}$Ca  & -72.5  &&  $^{44}$Ti & -91.2 \\ 
3  & $^{45}$V   & +56.6  &&  $^{57}$Co & +27.1 \\ 
4  & $^{36}$Ar  & -49.7  &&  $^{40}$Ca & +22.0 \\ 
5  & $^{57}$Ni  & -49.5  &&  $^{57}$Ni & +11.5 \\ 
6  & $^{57}$Co  & -33.0  &&  $^{54}$Fe & +9.4  \\ 
7  & $^{32}$S   & -20.4  &&  $^{36}$Ar & +5.6  \\ 
8  & $^{13}$N   & -14.8  &&  $^{46}$Cr & -5.3  \\ 
9  & $^{58}$Cu  & -13.5  &&  $^{56}$Ni & -5.2  \\ 
10 & $^{56}$Ni  & -4.5   &&  $^{61}$Cu & +4.0  \\ 
11 & $^{28}$Si  & -4.2   &&  $^{58}$Ni & +3.4  \\ 
12 & $^{56}$Co  & -4.1   &&  $^{23}$Na & -2.6  \\ 
13 & $^{20}$Ne  & -4.0   &&  $^{12}$C  & -1.8  \\ 
14 & $^{12}$C   & +3.1   &&  $^{13}$N  & -1.5  \\ 
15 & $^{48}$Cr  & +2.8   &&  $^{58}$Cu & +1.3  \\ 
16 & $^{39}$K   & -2.0   &&  $^{38}$Ca & +1.0  \\ 
17 & $^{14}$O   & +1.0   &&  $^{24}$Mg & +0.9  \\ 
18 & $^{27}$Al  & +0.9   &&  $^{59}$Cu & +0.9  \\ 
19 & $^{16}$O   & -0.8   &&  $^{32}$S  & +0.8  \\ 
20 & $^{24}$Mg  & -0.6   &&  $^{48}$Cr & -0.6  \\ 
\enddata 
\label{tab:tabasurv} 
\end{deluxetable}

\begin{deluxetable}{ccrccr} 
\tablecolumns{6} 
\tablewidth{0pc}
\tablecaption{
 Order of importance of reactions producing $^{44}$Ti at $\eta$=0 from low 
 multiplication factor (0.01$\times$) and high multiplication factor
 (100$\times$). 
} 
\tablehead{ 
\colhead{Rank} &
\multicolumn{2}{c}{Reaction rate} & \colhead{} &
\multicolumn{2}{c}{Reaction rate}  \\
     & \multicolumn{2}{c}{multiplied by 1/100} & &
       \multicolumn{2}{c}{multiplied by 100}  \\ 
\colhead{}  & \colhead{Reaction} & \colhead{\% $^{44}$Ti change} & \colhead{ } &
 \colhead{Reaction} & \colhead{\% $^{44}$Ti change} } 
\startdata 
1 & $^{44}$Ti($\alpha$,p)$^{47}$V           & +173 &&
    $^{45}$V(p,$\gamma$)$^{46}$Cr           & -98  \nl 
2 & $\alpha$(2$\alpha$,$\gamma$)$^{12}$C    & -100 &&
    $\alpha$(2$\alpha$,$\gamma$)$^{12}$C    & +67  \nl 
3 & $^{40}$Ca($\alpha$,$\gamma$)$^{44}$Ti   & -72  &&
    $^{44}$Ti($\alpha$,p)$^{47}$V           & -89  \nl 
4 & $^{45}$V(p,$\gamma$)$^{46}$Cr           & +57  &&
    $^{44}$Ti($\alpha$,$\gamma$)$^{48}$Cr   & -61  \nl 
5 & $^{57}$Ni(p,$\gamma$)$^{58}$Cu          & -47  &&
    $^{57}$Co(p,n)$^{57}$Ni                 & +25  \nl 
6 & $^{57}$Co(p,n)$^{57}$Ni                 & -33  &&
    $^{40}$Ca($\alpha$,$\gamma$)$^{44}$Ti   & +22  \nl 
7 & $^{13}$N(p,$\gamma$)$^{14}$O            & -16  &&
    $^{57}$Ni(n,$\gamma$)$^{58}$Ni          & +10  \nl 
8 & $^{58}$Cu(p,$\gamma$)$^{59}$Zn          & -14  &&
    $^{54}$Fe($\alpha$,n)$^{57}$Ni          & +9.4 \nl 
9 & $^{36}$Ar($\alpha$,p)$^{39}$K           & -11  &&
    $^{36}$Ar($\alpha$,p)$^{39}$K           & +5.5 \nl 
10 & $^{12}$C($\alpha$,$\gamma$)$^{16}$O    & +3.5 &&
     $^{36}$Ar($\alpha$,$\gamma$)$^{40}$Ca  & +5.3 \nl 
\enddata
\label{tab:rsurvlowhigh} 
\end{deluxetable}

\clearpage
\begin{deluxetable}{lr} 
\footnotesize 
\tablecaption{
 Order of Importance of Reactions Producing $^{44}$Ti at $\eta$ = 0
According to the Slope of X($^{44}$Ti) Near the Standard Cross Section 
\label{slopetab}} 
\tablewidth{0pt}
\tablehead{ \colhead{Reaction}       & \colhead{Slope} } 
\startdata
$^{44}$Ti($\alpha$,p)$^{47}$V             & -0.394 \nl
$\alpha$(2$\alpha$,$\gamma$)$^{12}$C      & +0.386 \nl
$^{45}$V(p,$\gamma$)$^{46}$Cr             & -0.361 \nl
$^{40}$Ca($\alpha$,$\gamma$)$^{44}$Ti     & +0.137 \nl
$^{57}$Co(p,n)$^{57}$Ni                   & +0.102 \nl
$^{36}$Ar($\alpha$,p)$^{39}$K             & +0.037 \nl
$^{44}$Ti($\alpha$,$\gamma$)$^{48}$Cr     & -0.024 \nl
$^{12}$C($\alpha$,$\gamma$)$^{16}$O       & -0.017 \nl
$^{57}$Ni(p,$\gamma$)$^{58}$Cu            & +0.013 \nl
$^{58}$Cu(p,$\gamma$)$^{59}$Zn            & +0.011 \nl
$^{36}$Ar($\alpha$,$\gamma$)$^{40}$Ca     & +0.008 \nl
$^{44}$Ti(p,$\gamma$)$^{45}$V             & -0.005 \nl
$^{57}$Co(p,$\gamma$)$^{58}$Ni            & +0.002 \nl
$^{57}$Ni(n,$\gamma$)$^{58}$Cu            & +0.002 \nl
$^{54}$Fe($\alpha$,n)$^{57}$Ni            & +0.002 \nl
$^{40}$Ca($\alpha$,p)$^{43}$Sc            & -0.002 \nl \enddata
\label{tab:tabslope} 
\end{deluxetable}

\clearpage
\begin{deluxetable}{ccrccr} 
\tablecolumns{6} 
\tablewidth{0pc}
\tablecaption{
 Order of importance of reactions producing $^{44}$Ti at $\eta$=0.002
 from low multiplication factor (0.01$\times$) and high multiplication 
 factor (100$\times$). 
} 
\tablehead{ 
\colhead{Rank} &
\multicolumn{2}{c}{Reaction rate} & \colhead{} &
\multicolumn{2}{c}{Reaction rate}  \\
     & \multicolumn{2}{c}{multiplied by 1/100} & &
       \multicolumn{2}{c}{multiplied by 100}  \\ 
\colhead{}  & \colhead{Reaction} & \colhead{\% $^{44}$Ti change} & \colhead{ } &
 \colhead{Reaction} & \colhead{\% $^{44}$Ti change} } 
\startdata 
1 & $^{44}$Ti($\alpha$,p)$^{47}$V           & +208 &&
    $^{44}$Ti($\alpha$,p)$^{47}$V           & -93  \nl
2 & $^{12}$C($\alpha$,$\gamma$)$^{16}$O     & -72 &&
    $^{44}$Ti($\alpha$,$\gamma$)$^{48}$Cr   & -66  \nl
3 & $^{40}$Ca($\alpha$,$\gamma$)$^{44}$Ti   & -66 &&
    $^{27}$Al($\alpha$,n)$^{30}$P           & -60 \nl
4 & $^{20}$Ne($\alpha$,$\gamma$)$^{24}$Mg   & -16 &&
    $^{30}$Si($\alpha$,n)$^{33}$S           & -33       \nl
5 & $^{30}$Si(p,$\gamma$)$^{31}$P           & -9.2 &&
    $^{12}$C($\alpha$,$\gamma$)$^{16}$O     & +18       \nl
6 & $^{36}$Ar($\alpha$,p)$^{39}$K           & -7.9 &&
    $^{40}$Ca($\alpha$,$\gamma$)$^{44}$Ti   & +15       \nl
7 & $^{59}$Ni(p,n)$^{59}$Cu                 & -4.7 &&
    $^{23}$Na($\alpha$,p)$^{26}$Mg          & -4.7       \nl
8 & $^{59}$Ni(p,$\gamma$)$^{60}$Cu          & -4.7 &&
    $^{39}$K($\alpha$,p)$^{42}$Ca           & +4.7       \nl
9 & $^{44}$Ti($\alpha$,$\gamma$)$^{48}$Cr   & +2.8 &&
    $^{27}$Al(p,$\gamma$)$^{28}$Si          & +4.3       \nl
10 & $^{27}$Al($\alpha$,n)$^{30}$P          & +2.7  &&
     $^{24}$Mg($\alpha$,$\gamma$)$^{28}$Si  & +4.2       \nl
\enddata
\label{tab:rsurvlowhigh2} 
\end{deluxetable}

\begin{deluxetable}{ccrccr} 
\tablecolumns{6} 
\tablewidth{0pc}
\tablecaption{
 Order of importance of reactions producing $^{44}$Ti at $\eta$=0.006
 from low multiplication factor (0.01$\times$) and high multiplication 
 factor (100$\times$). 
} 
\tablehead{ 
\colhead{Rank} &
\multicolumn{2}{c}{Reaction rate} & \colhead{} &
\multicolumn{2}{c}{Reaction rate}  \\
     & \multicolumn{2}{c}{multiplied by 1/100} & &
       \multicolumn{2}{c}{multiplied by 100}  \\ 
\colhead{}  & \colhead{Reaction} & \colhead{\% $^{44}$Ti change} & \colhead{ } &
 \colhead{Reaction} & \colhead{\% $^{44}$Ti change} } 
\startdata 
1 & $^{44}$Ti($\alpha$,p)$^{47}$V           & +211 &&
    $^{44}$Ti($\alpha$,p)$^{47}$V           & -93  \nl
2 & $^{12}$C($\alpha$,$\gamma$)$^{16}$O     & -79 &&
    $^{44}$Ti($\alpha$,$\gamma$)$^{48}$Cr   & -65  \nl
3 & $^{40}$Ca($\alpha$,$\gamma$)$^{44}$Ti   & -65 &&
    $^{27}$Al($\alpha$,n)$^{30}$P           & -56 \nl
4 & $^{20}$Ne($\alpha$,$\gamma$)$^{24}$Mg   & -11 &&
    $^{30}$Si($\alpha$,n)$^{33}$S           & -39       \nl
5 & $^{30}$Si(p,$\gamma$)$^{31}$P           & -9.6 &&
    $^{12}$C($\alpha$,$\gamma$)$^{16}$O     & +19       \nl
6 & $^{36}$Ar($\alpha$,p)$^{39}$K           & -7.5 &&
    $^{40}$Ca($\alpha$,$\gamma$)$^{44}$Ti   & +15       \nl
7 & $^{27}$Al($\alpha$,p)$^{30}$Si          & -4.0  &&
    $^{58}$Ni($\alpha$,$\gamma$)$^{62}$Zn   & -8.7       \nl
8 & $^{33}$S(p,$\gamma$)$^{34}$Cl           & +3.8 &&
    $^{27}$Al(p,$\gamma$)$^{28}$Si          & +6.0       \nl
9 & $^{16}$O($\alpha$,$\gamma$)$^{20}$Ne    & -3.8 &&
     $^{24}$Mg($\alpha$,$\gamma$)$^{28}$Si  & +6.0       \nl
10 & $^{30}$Si($\alpha$,n)$^{33}$S          & +3.5  &&
    $^{39}$K($\alpha$,p)$^{42}$Ca           & +5.3       \nl
\enddata
\label{tab:rsurvlowhigh6} 
\end{deluxetable}

\clearpage
\begin{deluxetable}{lrrrrrrr} 
\footnotesize 
\tablecaption{\bf 
 Fits to thermally-averaged cross sections $N_{A} <\sigma v>$ 
 (in cm$^3$ s$^{-1}$ mole$^{-1}$) used in our calculations
} 
\tablewidth{0pt}
\tablehead{ 
\colhead{Reaction}       & 
\colhead{A} & 
\colhead{B} &
\colhead{C} & 
\colhead{D} & 
\colhead{E} & 
\colhead{F} & 
\colhead{G} }
\startdata 
$^{44}$Ti($\alpha$,p)$^{47}$V &   110.00000 &    -3.84890
&   -47.13600 &   -91.31700 &     7.65930 &     -.54662 & 30.27300\nl 
$\alpha$(2$\alpha$,$\gamma$)$^{12}$C & CF88\nl
$^{45}$V(p,$\gamma$)$^{46}$Cr &    90.47900 &      .55899 &   
-37.67100 & -57.76700 &     5.08410 &     -.31734 &    10.46700\nl
$^{40}$Ca($\alpha$,$\gamma$)$^{44}$Ti &   119.01000 & -4.29580 &
105.93000 &  -250.83000 &    10.96000 &     -.50058 & 128.14999\nl
$^{57}$Co(p,n)$^{57}$Ni &    13.38000 &   -46.93500 &    -3.84000 &
9.20780 &     -.27549 &      .00430 &    -3.93210\nl
$^{36}$Ar($\alpha$,p)$^{39}$K &    63.29300 &   -14.78800 &   -29.48900
&   -42.76300 &     5.14390 &     -.42968 &    11.34400\nl
$^{44}$Ti($\alpha$,$\gamma$)$^{48}$Cr &    81.02500 &     4.26030 &
-245.58000 &   137.92000 &    -2.79730 &     -.07990 &   -98.51900\nl
$^{12}$C($\alpha$,$\gamma$)$^{16}$O & $1.7 \times$ CF88\nl
$^{57}$Ni(p,$\gamma$)$^{58}$Cu &   109.94000 &    -1.16540 &
35.84500 &  -155.78999 &     9.22410 &     -.50439 &    64.77700\nl
$^{58}$Cu(p,$\gamma$)$^{59}$Zn &   118.27000 &    -1.34030 &
45.34600 &  -175.61000 &    10.46300 &     -.59085 &    73.47000\nl
$^{36}$Ar($\alpha$,$\gamma$)$^{40}$Ca &   273.35001 &    -5.81920 &
235.08000 &  -552.10999 &    34.32800 &    -2.01860 &   245.53999\nl
$^{44}$Ti(p,$\gamma$)$^{45}$V &    81.78900 &      .07713 &   -30.83200
&   -55.66900 &     5.66600 &     -.43928 &    10.56000\nl
$^{57}$Co(p,$\gamma$)$^{58}$Ni &    70.29000 &    -1.41470 &
30.19200 &  -105.66000 &     5.11650 &     -.27513 &    50.64400\nl
$^{57}$Ni(n,$\gamma$)$^{58}$Cu &    15.67600 &     -.01193 &
.42180 &     -.30912 &      .08421 &     -.01823 &      .01975\nl
$^{54}$Fe($\alpha$,n)$^{57}$Ni &    17.59100 &   -67.55000 &
1.68610 &    -4.90450 &      .88835 &     -.05892 &     1.50260\nl
$^{40}$Ca($\alpha$,p)$^{43}$Sc &    52.60400 &   -40.91600 &
-22.86600 &   -28.98800 &     1.59900 &     -.06980 &    10.16200\nl
\enddata 
\tablenotetext{}{
 The rates for these reactions are from CF88 (\cite{CF88}) or are from 
 the formula 
 $N_A < \sigma v > = \exp(A + B/T_9 + C/T_9^{1/3} + D T_9^{1/3} + E T_9 
 + F T_9^{5/3} + G \ln T_9)$.}
\label{tab:ourrates} 
\end{deluxetable}



\clearpage 
\setcounter{figure}{0}
\begin{figure} 
\plotfiddle{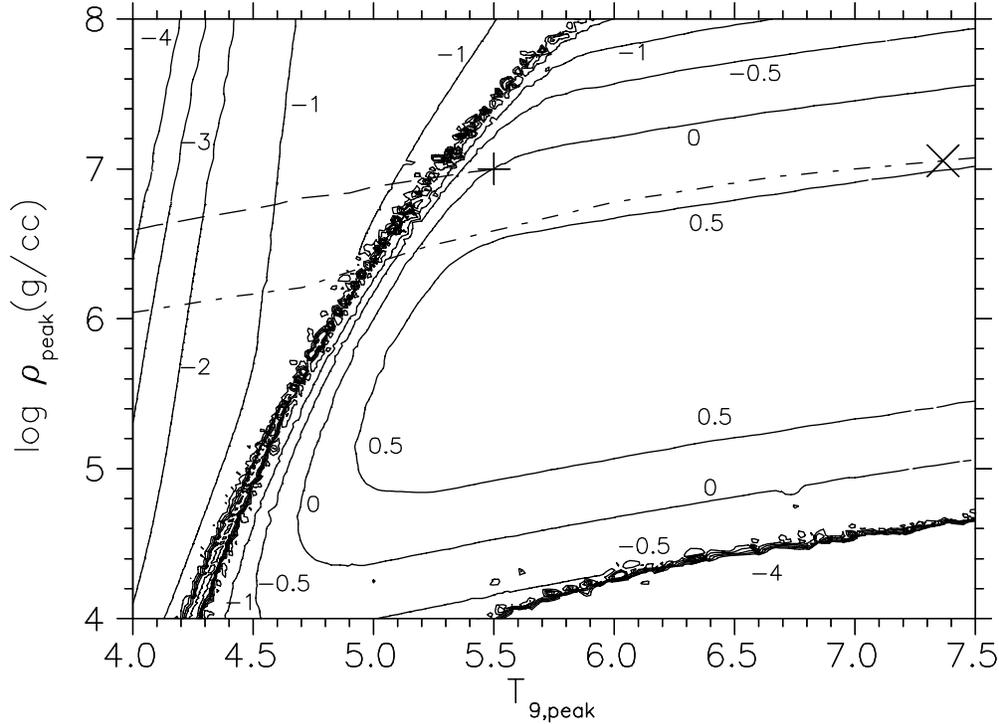}{240pt}{90}{70}{70}{240}{-80}
\vspace{0.5in}
\caption{ 
 Final $^{44}$Ti mass fraction in a zone expanding with
 $\rho\propto T^3$ as a function of initial density (g/cc) and
 temperature (K) for initial composition consisting of pure $^{28}$Si.
 The contour levels show the logarithm of the final $^{44}$Ti mass
 fractions relative to the final $^{44}$Ti mass fraction in our
 reference calculation that began with initial temperature T$_9$=5.5 and
 density $\rho$=10$^7$ g/cc (+ sign).  This figure shows that the
 initial temperature and density chosen for the survey in the rest of
 this paper (+ sign) produces a final $^{44}$Ti mass fraction that is
 roughly equal to the final $^{44}$Ti mass fraction over a quite large
 range of initial temperatures and densities.  Most of the plane is
 within a factor 10.  The dashed line shows the temporal evolution of
 temperature and density in the $\rho\propto T^3$ expansion of our
 survey.  The dash-dotted line shows the peak temperatures and densities
 of matter in the ejecta of the 20 M$_{\sun}$ supernova model of
 Thielemann et al. (1990).  The $\times$ sign locates the mass cut of
 the model at $\sim$1.6 M$_{\sun}$ radius in order to eject 0.07
 M$_{\sun}$ of $^{56}$Ni.  It is important to note that unlike the
 dashed curve, the dash-dotted curve is not a time history of a single
 zone. Rather, it shows the peak conditions of $\rho$ and T$_9$ reached
 by a continuum of mass zones in the explosion. The conditions chosen
 for our reference calculation are reasonably close to those achieved in
 supernova models.  The ridge extending from the bottom left to the top
 middle of the figure comprises initial conditions for which the final
 $^{44}$Ti mass fraction reaches its highest yield; this is the
 condition where the abundance of heavy nuclei within QSE  is much the
 same as the abundance of heavy nuclei within NSE from the moment when
 quasi-equilibrium  is first achieved until the freezeout.
 \label{fig:Ti44contour} 
} 
\end{figure}
\clearpage

\begin{figure} 
\plotfiddle{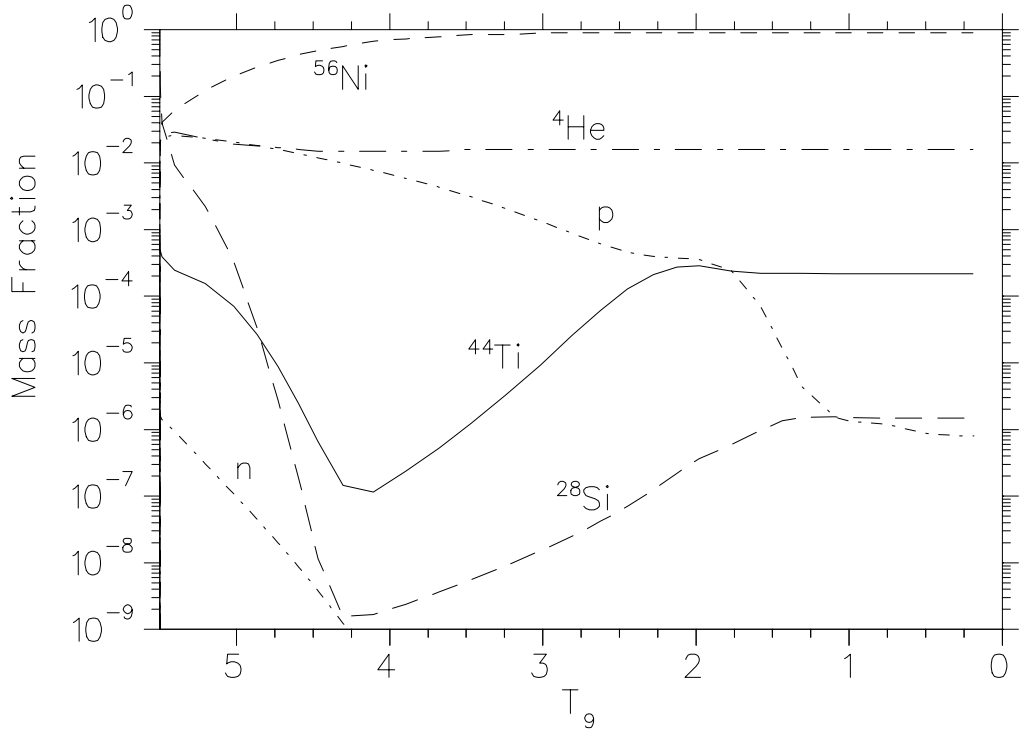}{240pt}{90}{70}{70}{240}{-80}
\vspace{0.5in} 
\caption{ 
 The evolution of the mass fractions of some important nuclei in the 
 nearly adiabatic expansion experiencing alpha-rich freezeout with 
 standard nuclear reaction rates and initial temperature T$_9$=5.5 
 and density $\rho$=10$^7$ g/cc.  \label{fig:massfr} 
} 
\end{figure}
\clearpage

\begin{figure} 
\plotfiddle{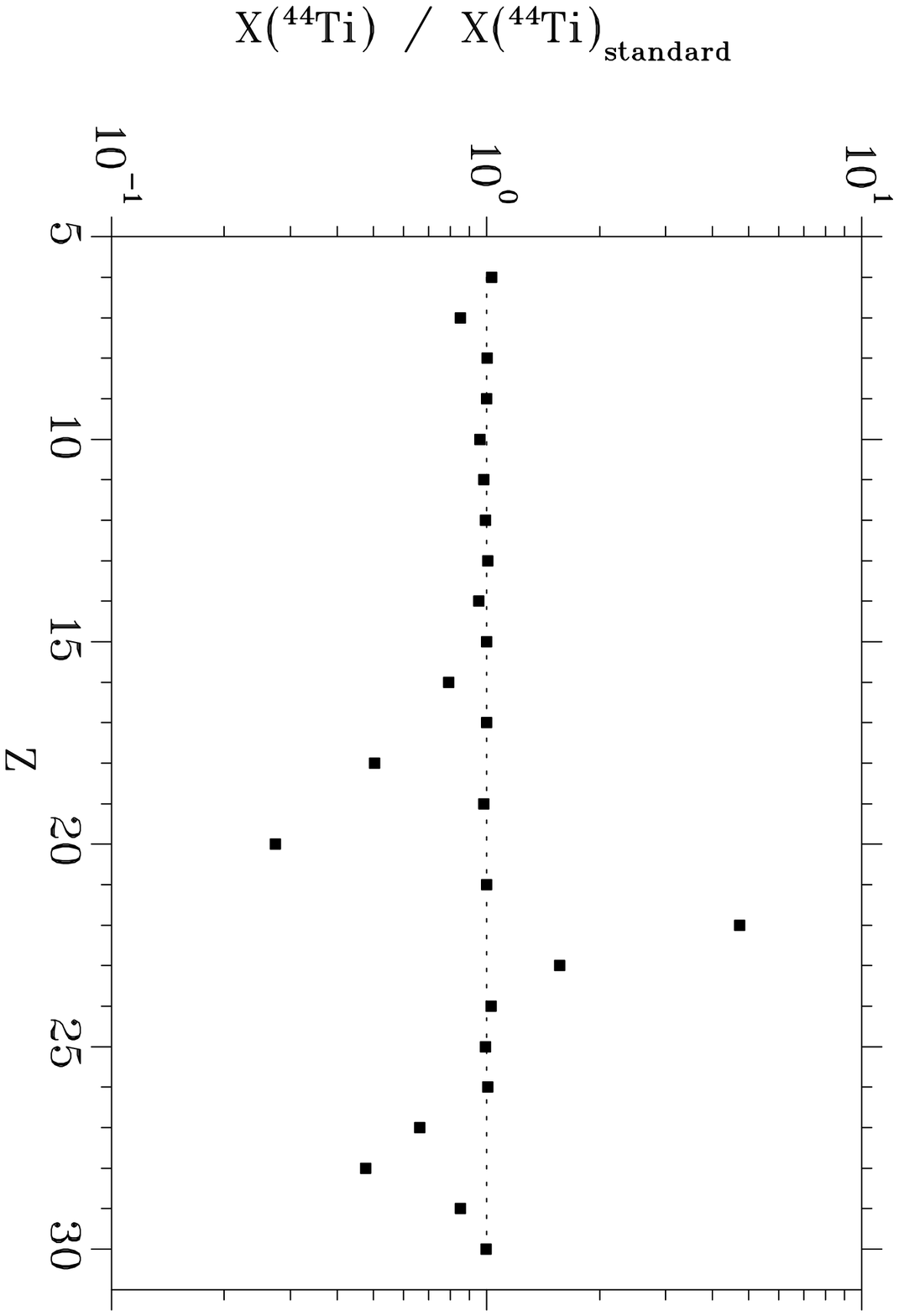}{220pt}{90}{50}{50}{200}{-80} 
\vspace{0.6in}
\plotfiddle{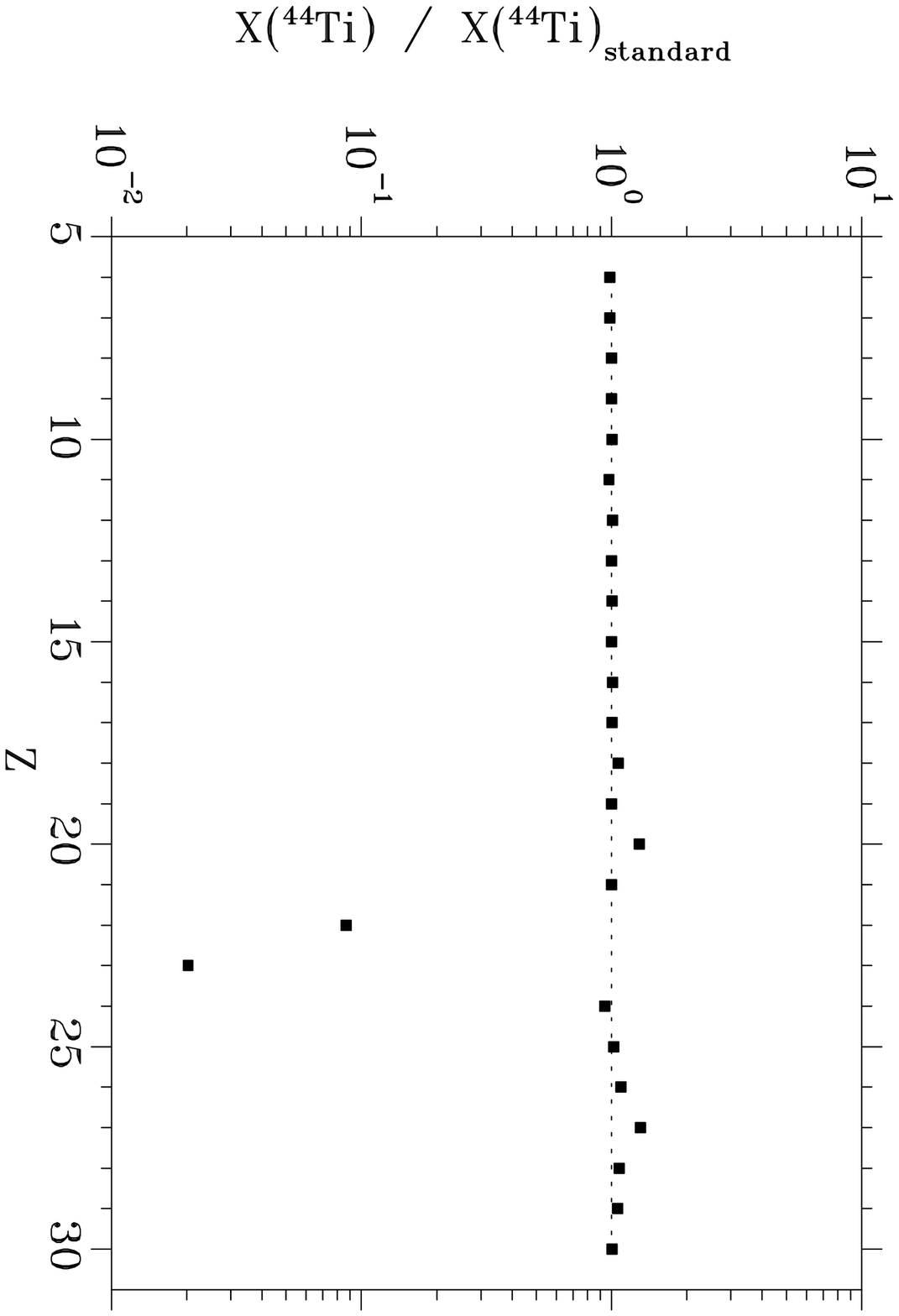}{220pt}{90}{50}{50}{200}{-80} 
\vspace{0.6in}
\caption{ 
 The ratio of the final $^{44}$Ti mass fraction to the $^{44}$Ti mass 
 fraction from the standard network if the reaction rates of all 
 charge-increasing and mass-increasing reactions 
 of all isotopes of a single element are multiplied, element by element, 
 by 0.01(a) or 100 (b). \label{fig:zsurv} 
}
\end{figure}

\begin{figure} 
\plotfiddle{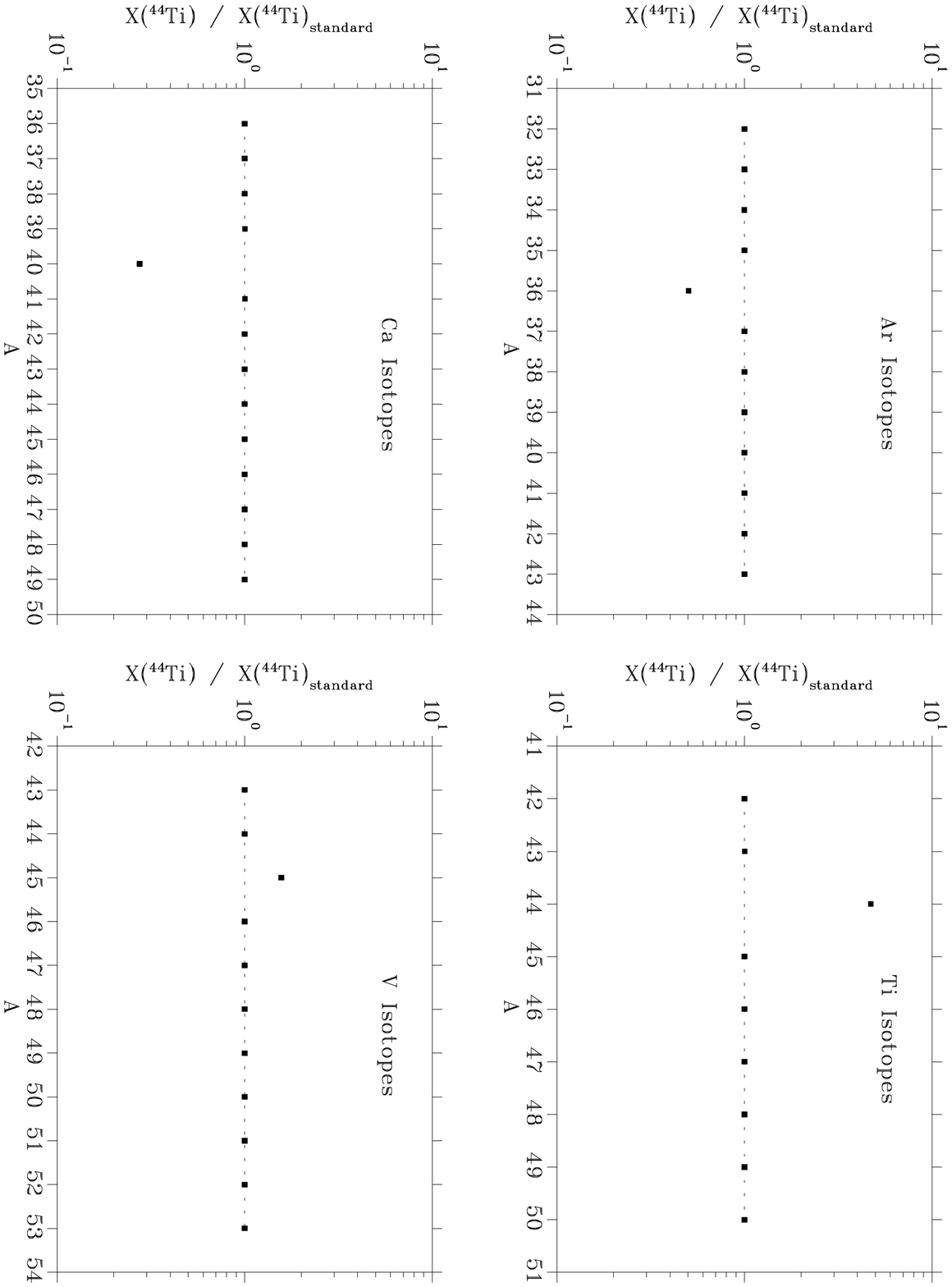}{220pt}{90}{50}{50}{190}{-60}
\vspace{0.7in} 
\plotfiddle{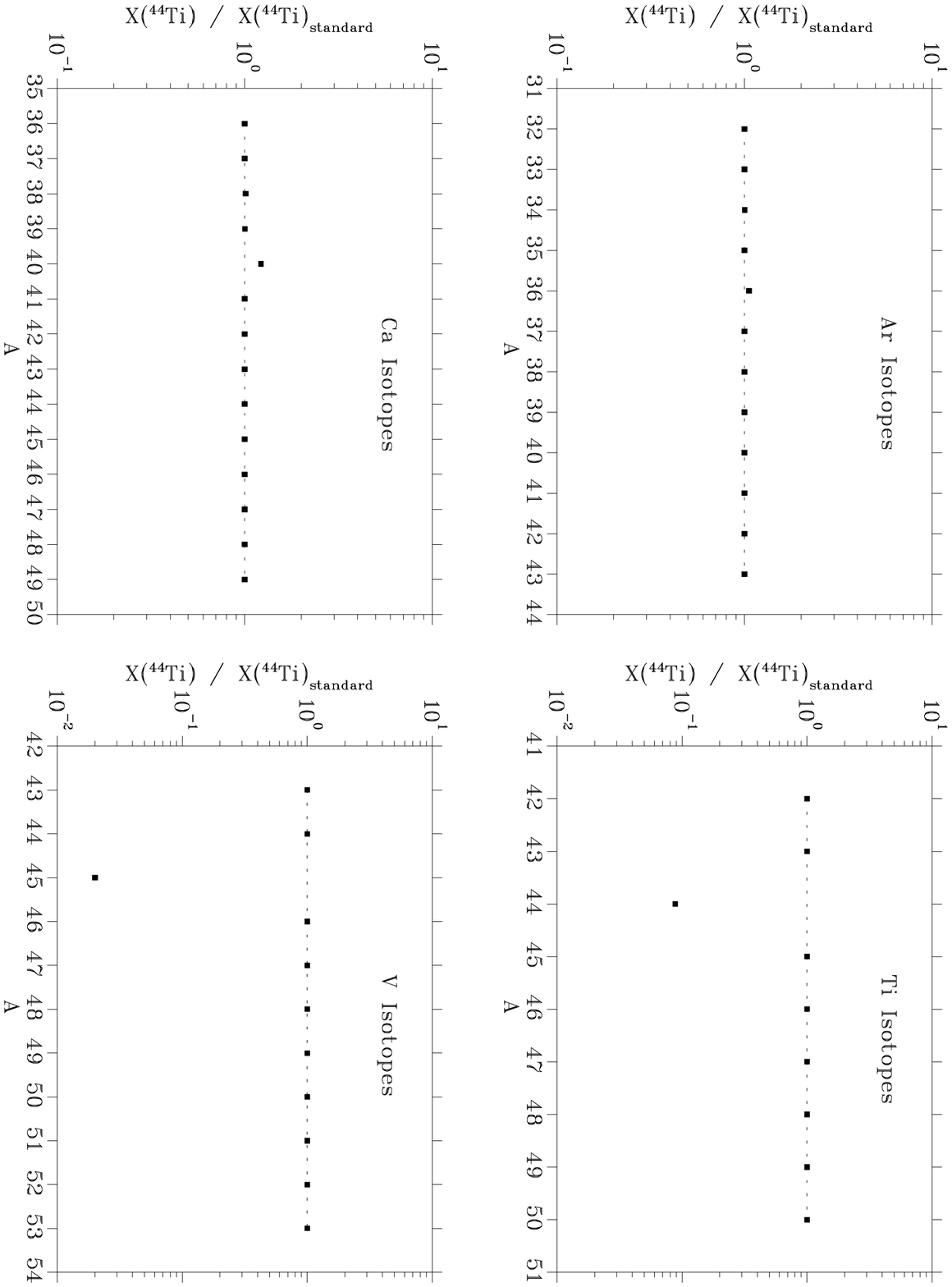}{220pt}{90}{50}{50}{190}{-80}
\vspace{0.7in} 
\caption{ 
 X($^{44}$Ti)/X($^{44}$Ti)$_{standard}$ for isotopic reaction rates 
 decreased by $\times$0.01 (a) and increased by $\times$100 (b).  
 Reactions with $^{44}$Ti and $^{45}$V produce the largest abundance 
 sensitivity. \label{fig:asurv} 
} 
\end{figure}

\begin{figure} 
\plotfiddle{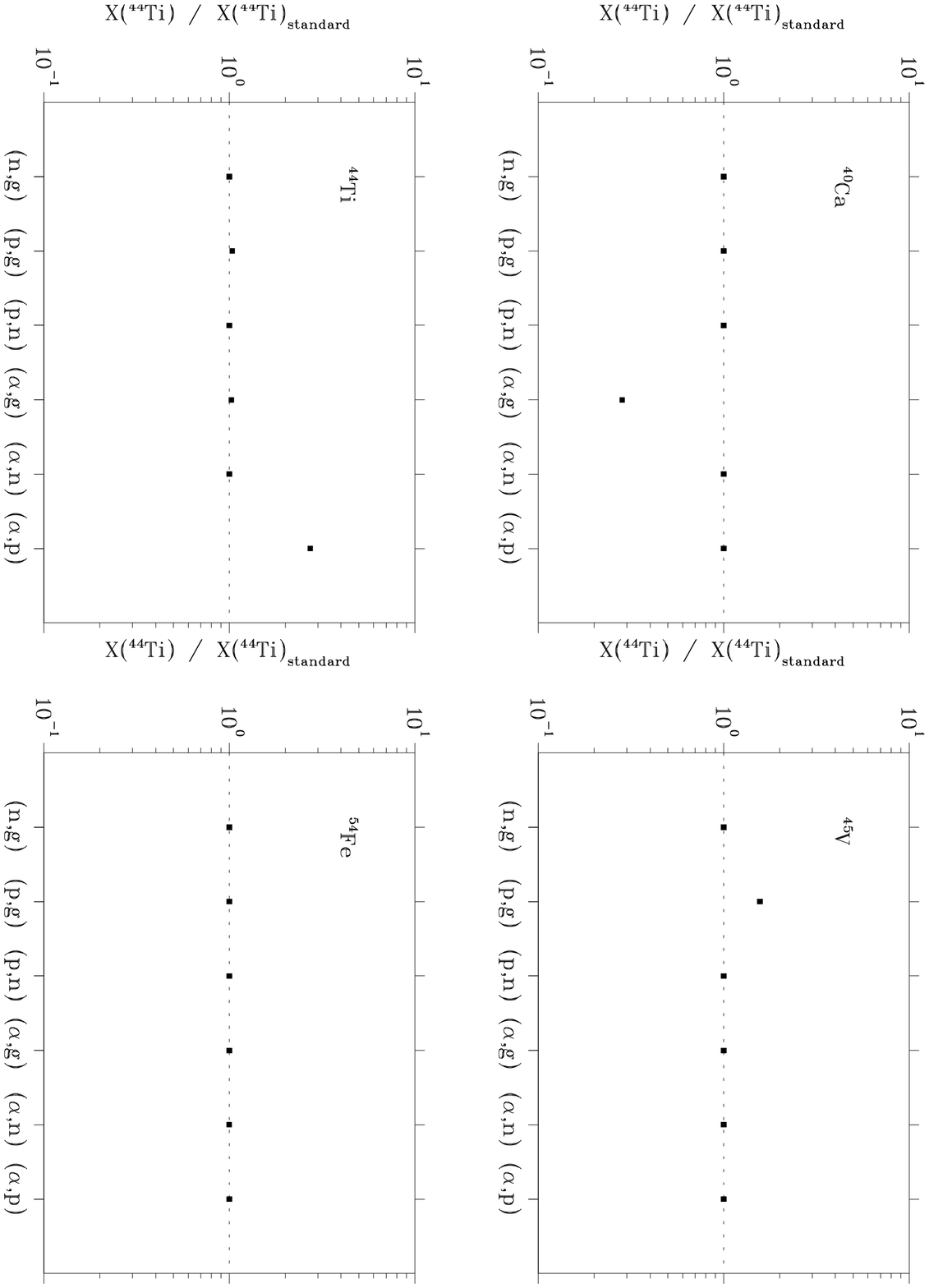}{220pt}{90}{50}{50}{190}{-60}
\vspace{0.7in} 
\plotfiddle{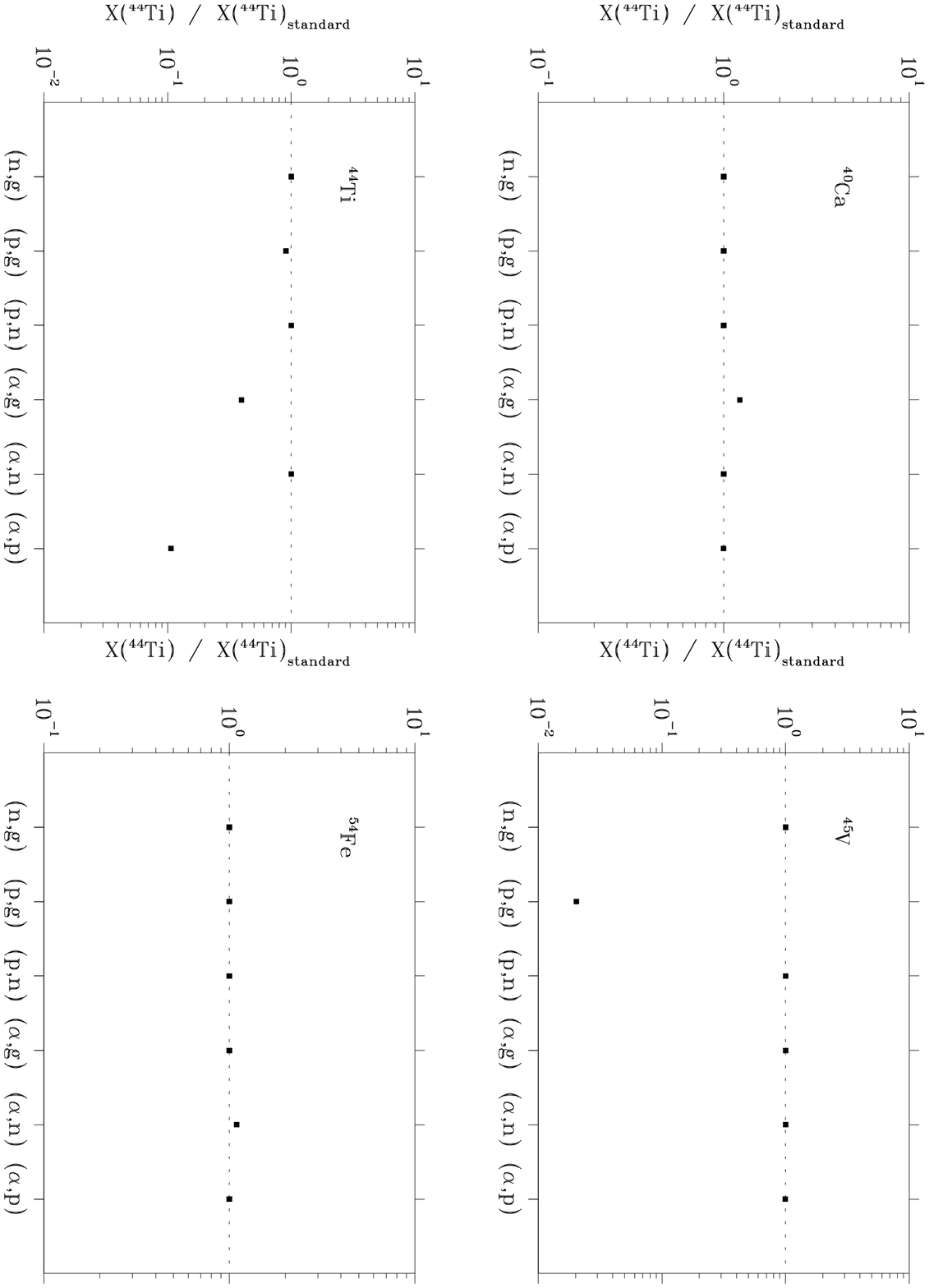}{220pt}{90}{50}{50}{190}{-60}
\vspace{0.7in} 
\caption{ 
 X($^{44}$Ti)/X($^{44}$Ti)$_{standard}$ for reaction rates 
 decreased by $\times$0.01 (top) and increased by $\times$100 (bottom).  
 Reactions with $^{44}$Ti and $^{45}$V produce the largest abundance 
 sensitivity. \label{fig:rsurv} 
} 
\end{figure}

\begin{figure} 
\plotfiddle{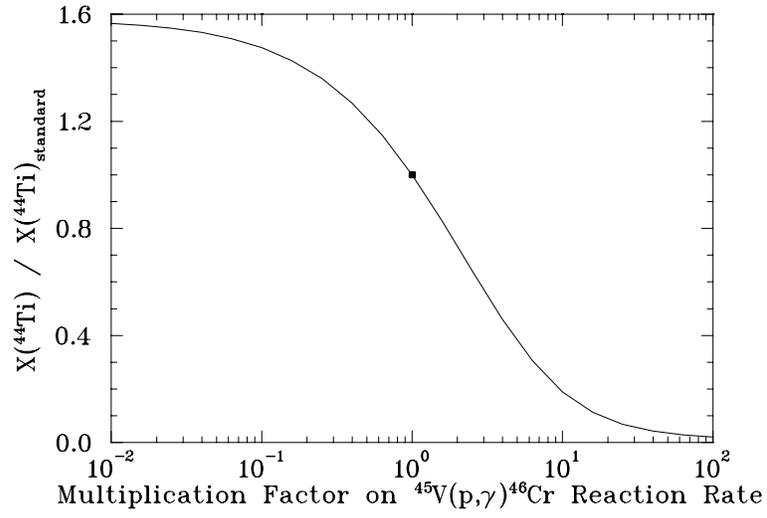}{220pt}{90}{50}{50}{200}{-80}
\vspace{0.4in} 
\caption{ 
 Final X($^{44}$Ti) dependence on the reaction rate of 
 $^{45}$V(p,$\gamma$)$^{46}$Cr.  \label{fig:45pg} 
} 
\end{figure}

\begin{figure} 
\plotfiddle{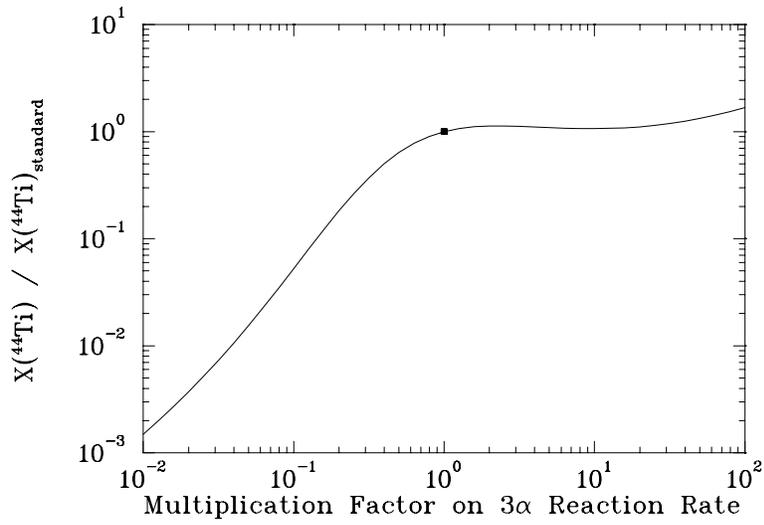}{220pt}{90}{50}{50}{200}{-80}
\vspace{0.4in} 
\caption{ 
 Final X($^{44}$Ti) dependence on the reaction rate of 
 $\alpha$(2$\alpha$).  \label{fig:3alpha} 
} 
\end{figure}

\begin{figure} 
\plotfiddle{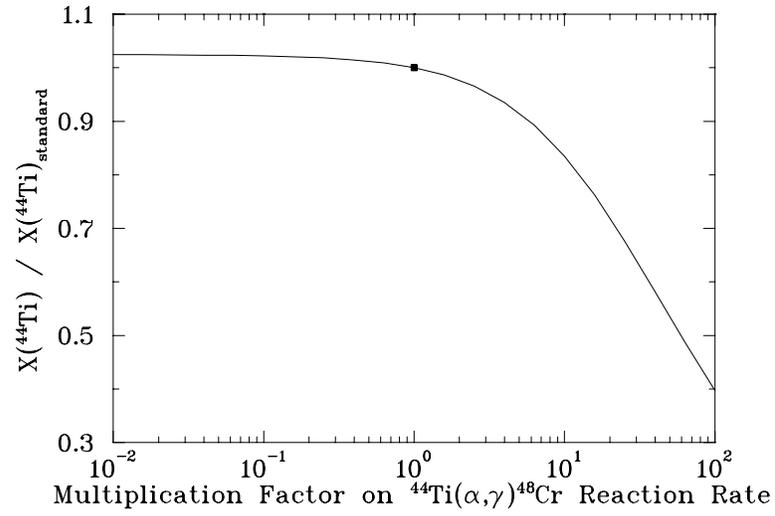}{220pt}{90}{50}{50}{200}{-80}
\vspace{0.4in} 
\caption{ 
 Final X($^{44}$Ti) dependence on the reaction rate of 
 $^{44}$Ti($\alpha$,$\gamma$)$^{48}$Cr.  \label{fig:44ag} 
}
\end{figure}

\begin{figure} 
\plotfiddle{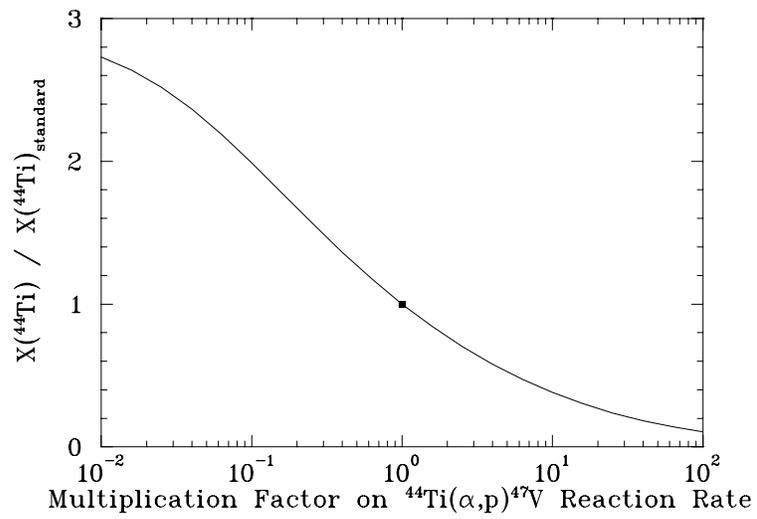}{220pt}{90}{50}{50}{200}{-80}
\vspace{0.4in} 
\caption{ 
 Final X($^{44}$Ti) dependence on the reaction rate of 
 $^{44}$Ti($\alpha$,p)$^{47}$V. \label{fig:44ap} 
}
\end{figure}

\begin{figure}[t] 
\plotfiddle{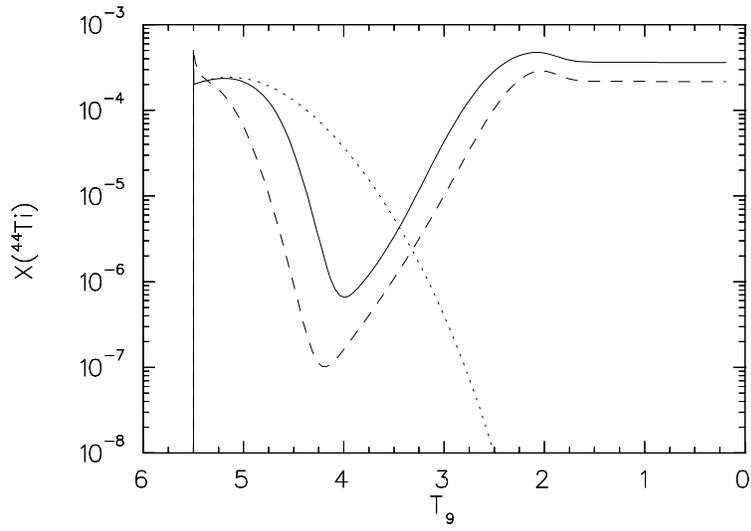}{220pt}{90}{50}{50}{200}{-80}
\vspace{0.4in} 
\caption{ 
 Evolution of the $^{44}$Ti mass fraction as the temperature falls, 
 with the standard triple-alpha rate (dashed curve) and also with the 
 rate increased by $\times$100 (solid line).
 The dotted line shows the NSE mass fraction of $^{44}$Ti during the
 expansion.  \label{fig:triple-a} 
} 
\end{figure}

\begin{figure}[t] 
\plotfiddle{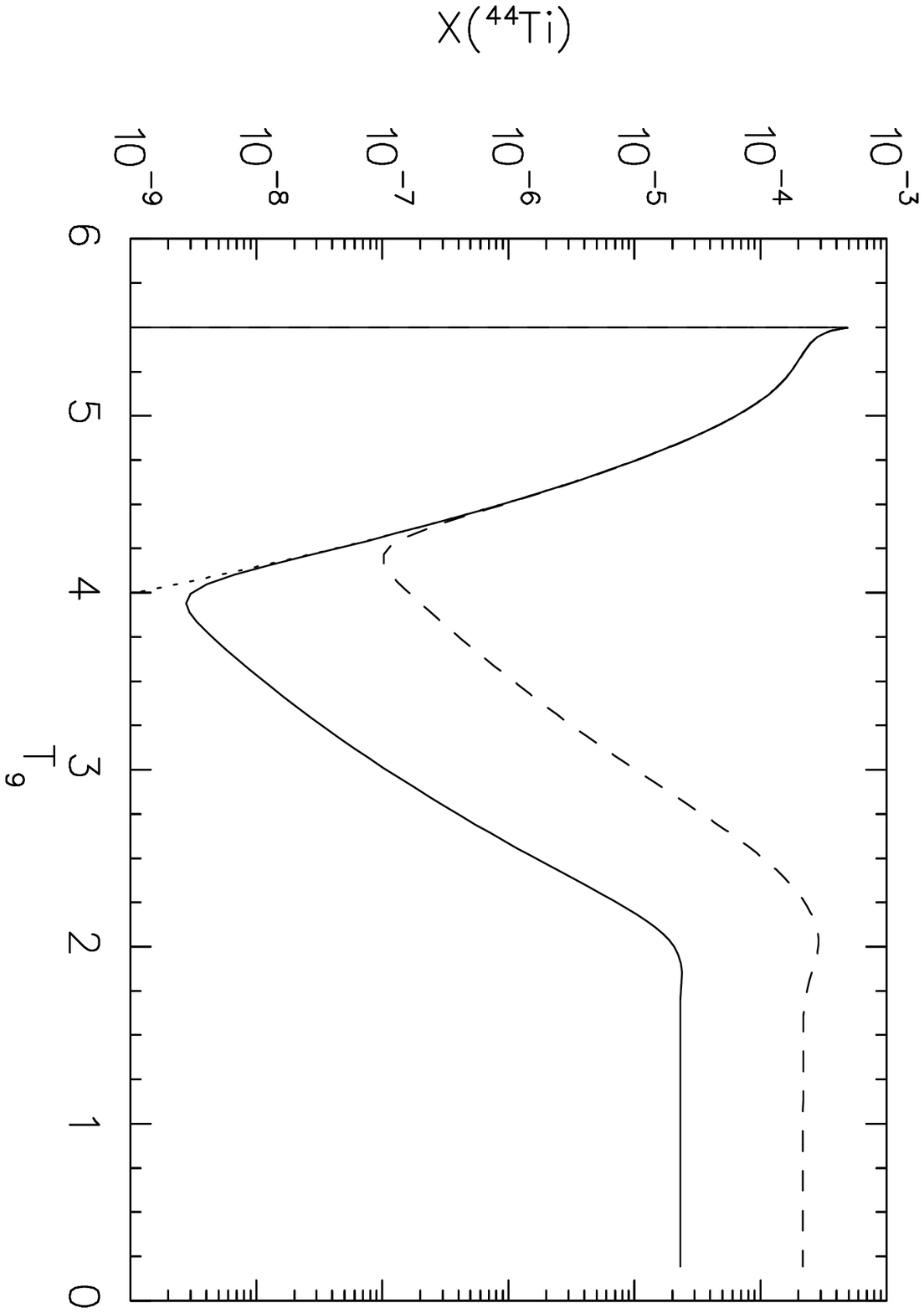}{220pt}{90}{50}{50}{200}{-80}
\vspace{0.4in} 
\caption{ 
 Evolution of the $^{44}$Ti mass fraction as the temperature falls, 
 with the standard $^{44}{\rm Ti}(\alpha,p)^{47}{\rm V}$ rate 
 (dashed curve) and also with the rate increased by 
 $\times$100 (solid line).  The dotted line shows the QSE
 mass fraction of $^{44}$Ti during the expansion.  \label{fig:ti44ap} 
}
\end{figure}

\begin{figure}[t] 
\plotfiddle{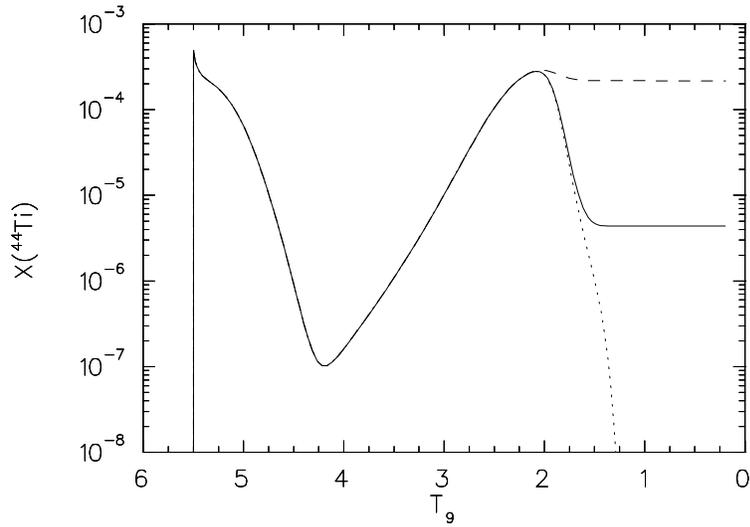}{220pt}{90}{50}{50}{200}{-80} 
\vspace{0.4in} 
\caption{ 
 Evolution of the $^{44}$Ti mass fraction as
 the temperature falls, with the standard 
 $^{45}{\rm V}(p,\gamma)^{46}{\rm Cr}$ rate (dashed curve) and 
 also with the rate increased by $\times$100 (solid line).  
 The dotted line shows the mass fraction of $^{44}$Ti during the 
 expansion if $(p,\gamma)-(\gamma,p)$ equilibrium for the $N=22$ isotones 
 were able to persist.  \label{fig:v45pg} 
} 
\end{figure}

\begin{figure}[t] 
\plotfiddle{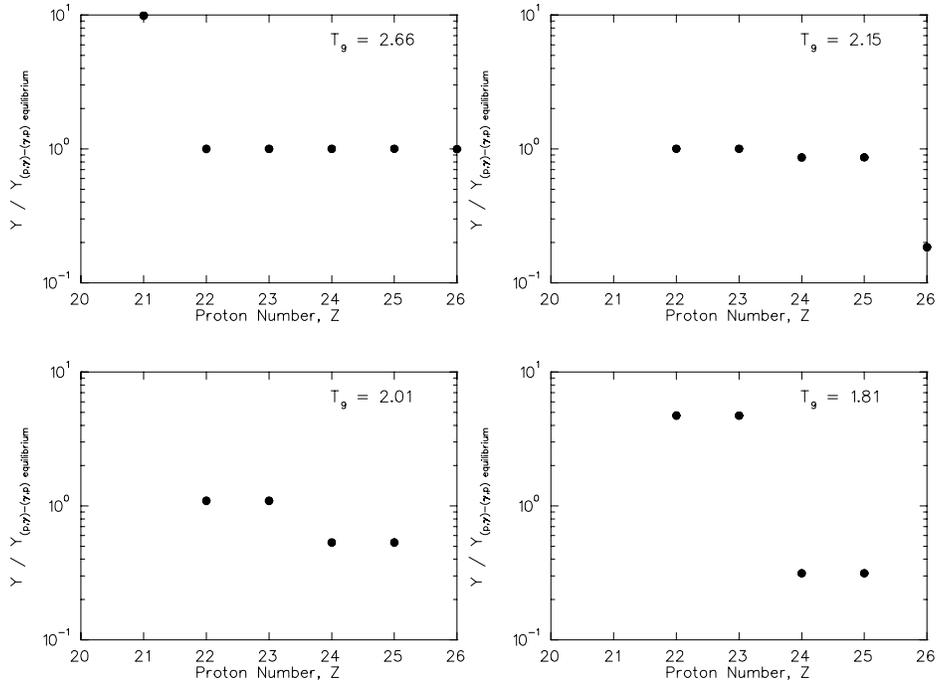}{220pt}{90}{50}{50}{200}{-80}
\vspace{0.8in} 
\caption{ 
 Evolution of the $(p,\gamma)-(\gamma,p)$ equilibrium for the N=22 
 isotones as the temperature falls, with the standard 
 $^{45}$V(p,$\gamma$)$^{46}$Cr rate.  
 The $^{44}$Ti and $^{45}$V pair (Z=22 and 23) increasingly depart 
 from equilibrium with the $^{46}$Cr and $^{47}$Mn pair (Z=24 and 25) 
 as the temperature declines.  \label{fig:pgequil} 
} 
\end{figure}

\begin{figure} 
\plotfiddle{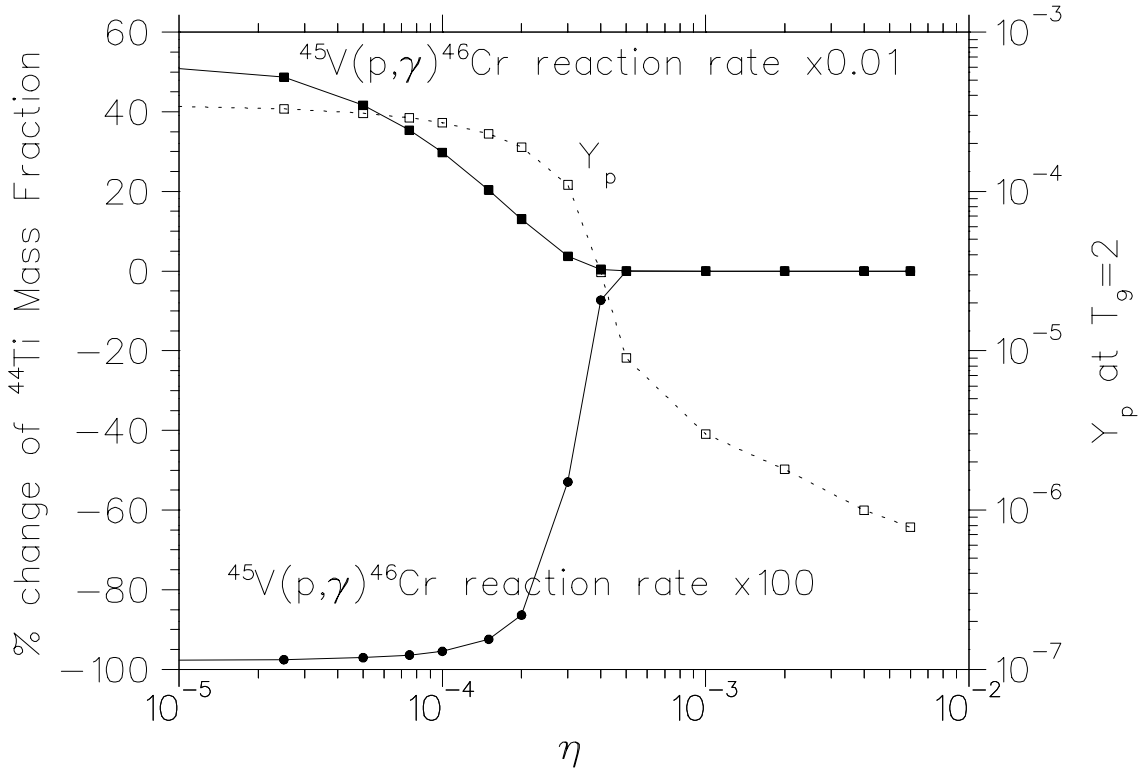}{240pt}{90}{70}{70}{240}{-80}
\vspace{0.5in} 
\caption{ 
 The percentage changes of $^{44}$Ti mass fraction as a function of $\eta$
 when the standard $^{45}$V(p,$\gamma$)$^{46}$Cr reaction rate is decreased 
 by $\times$0.01 (filled squares) and increased by $\times$100 (filled
 circles). The importance of $^{45}$V(p,$\gamma$)$^{46}$Cr reaction rate 
 to the $^{44}$Ti production in alpha-rich freezeout condition
 declines at $\eta \geq$ 0.0004.
 The dotted line (open squares) shows the proton abundance at
 T$_9$=2 as a function of $\eta$.  
 The drop of the proton abundance at $\eta \geq 0.0004$ causes the 
 diminishing importance of $^{45}$V(p,$\gamma$)$^{46}$Cr reaction rate 
 for increasing neutron richness.
 \label{fig:V45pgeta} 
} 
\end{figure}
\clearpage

\end{document}